\documentclass[twocolumn,a4paper,showpacs,aps,pra] {revtex4-1}

\usepackage{graphicx,color}
\graphicspath{{figure/}}

\begin{document}

\title{Quantum-beat Auger spectroscopy}
\author {Song Bin Zhang$^{1,2,*,**}$ and Nina Rohringer$^{1,2,\dagger}$}
\affiliation {$^1$Max Planck Institute for the Structure and Dynamics of Matter, 22761 Hamburg, Germany\\
              $^2$Center for Free-Electron Laser Science (CFEL), 22761 Hamburg, Germany\\
              $^*$song-bin.zhang@mpsd.mpg.de \\
              $^\dagger$nina.rohringer@mpsd.mpg.de \\                          
              $^{**}$Present address: School of Physics and Information Technology, Shaanxi Normal University, Xi'an 710062, China}

\begin{abstract}
The concept of nonlinear quantum-beat pump-probe Auger spectroscopy is introduced by discussing a relatively simple four-level model system. We consider a coherent wave packet involving two low-lying states that was prepared by an appropriate pump pulse. This wave packet is subsequently probed by a weak, time-delayed probe pulse with nearly resonant coupling to a core-excited state of the atomic or molecular system. The resonant Auger spectra are then studied as a function of the duration of the probe pulse and the time delay. With a bandwidth of the probe pulse approaching the energy spread of the wave packet, the Auger yields and spectra show quantum beats as a function of pump-probe delay. An analytic theory for the quantum-beat Auger spectroscopy will be presented, which allows for the reconstruction of the wave packet by analyzing the delay-dependent Auger spectra. The possibility of extending this method to a more complex manifold of electronic and vibrational energy levels is also discussed.
\end{abstract}

\maketitle

\section{Introduction}
In 1925, P. Auger discovered and interpreted the nonradiative decay process of an inner-shell ionized atom by the emission of an electron, which is nowadays known as Auger effect. In the following year, Robinson and Cassie investigated the Auger electrons by means of a magnetic electron spectrograph, which marks the birth date of Auger spectroscopy {\color{blue} \cite{Mehlhorn1998}}. After its discovery, Auger spectroscopy has been broadly applied and extensively developed in many different areas {\color{blue} \cite{chang1971,weissmann1981,moretti1998,Mehlhorn1998,carlson2013}}. The appearance of dedicated storage-ring x-ray radiation sources in the last quarter century has brought the Auger spectroscopy into a new period. With the rapid development of x-ray free electron laser pulses (FELs) in the past decade, the Auger spectroscopy
is expected to be pushed into a new level of application to study ultrafast electronic processes, considering the unique properties of FELs that are ultrashort pulses and ultrastrong intensities  {\color{blue} \cite{Ullrich2012}}. Actually, the development of FELs has spurred an increasing number of new studies of Auger spectroscopy in the soft x-ray region in atomic and molecular systems {\color{blue}\cite{Rohringer2008,liu2010,Kanter2011,Cederbaum2011a,Demekhin2011a,Demekhin2011b,Rohringer2012,Demekhin2013a,muller2015,Chatterjee2015,McFarland2014}}. 
In most of these studies, all kind of nonlinear interactions of the FELs on core-valence transitions are addressed, leading to a manipulated Auger spectrum in the strong-field limit. Contrary to those studies taking advantage of the ultrahigh intensities of FELs, we focus on the application of ultrashort FEl pulses with appropriate bandwidth and propose quantum-beat Auger spectroscopy to study coherent wave packet dynamics. We find that quantum-beat Auger spectroscopy, a linear probe process in the photon interaction, can be efficiently employed to extract the information of the initial coherent wave packet. \\

Many aspects have inspired us to propose the quantum-beat Auger spectroscopy. One scientific interest is the reconstruction of coherent electronic wave packets prepared by appropriate pulses. Various techniques {\color{blue}\cite{ Yudin2005,Yudin2006,loh2007,holler2011,Bredtmann2011,kraus2013,liu2014,liu2015}} have been proposed for the complete reconstruction of coherent wave packets in the optical or ultra violet regimes. Among those methods, it is worth mentioning that femto- and attosecond transient-absorption spectroscopy {\color{blue} \cite{loh2007,holler2011}} has become a very powerful one with the development of femtosecond and attosecond laser pulses, especially its successful application to extract the phase and amplitudes of laser induced wave packets {\color{blue}\cite{loh2007,Loh2008,Goulielmakis2010,ott2013,Chini2013,kaldun2014,liu2015}}. Transient absorption measurement, however, require optically dense samples and interpretation is sometimes cumbersome due to propagation effects of the applied laser pulses. A way to circumvent this problem, are detection of electrons rather than photons in optically thin samples. Read-out of spectroscopic information via the electron emission can also be an advantage in the development of nonlinear coherent x-ray pump probe techniques, based on stimulated x-ray Raman scattering  {\color{blue}\cite{schweigert2007,harbola2009}}. The transfer of these nonlinear optical techniques to the x-ray regime is not only challenged by the relatively low nonlinear optical susceptibilities in the x-ray region, beam propagation effects, and the relative unstable shot-to-shot properties of FELs. A potentially limiting issue in the x-ray regime is the direct coupling of the probe pulse to the ionization continuum, creating a large amount of ions in the sample. These ions can have resonances in the spectral region of interest of the nonlinear signal, and therefore can induce background, i.e. resonance absorption in homodyne detection schemes overlapping with typically low Raman signals {\color{blue}\cite{Rohringe2015}}. One way to circumvent these problems would be to read-out the coherent nonlinear response by Auger-electron spectroscopy, rather than detecting the signal photons or photo ionized electrons. Auger spectroscopy is proposed as a potential and complementary method in the x-ray regime to reconstruct the coherent wave packet. Here, we focus on one of the conceptually easiest nonlinear pump-probe techniques, quantum beat spectroscopy.\\

We present the method of quantum-beat Auger spectroscopy, by discussing a relatively simple four-level model system, shown in Fig.\  {\color{blue} 1}. In our analysis, we do  not discuss the process of creating a coherent wavepacket, but suppose a coherent wave packet, involving two bound states $|B_1\rangle$ and $|B_2\rangle$, that was created by an appropriate pump pulse, with the populations $c_1^2$ and $c_2^2$ ($c_2^2=1-c_1^2$), respectively, and a relative phase $\phi_0$. A weak femtosecond x-ray probe pulse triggering the Auger process is time delayed by $\Delta t$ from the pump pulse, and has a enter frequency $\omega_0$ between the resonant transition energies $E_{CB_1}$ and $E_{CB_2}$. In our notation $E_{I}$ refers to the energy of a particular state $|I\rangle$, and the energy differences between states $|J\rangle$ and $|I\rangle$are defined by $E_{JI}=E_J-E_I$. The femtosecond x-ray pulse prompts the wave packet to the intermediate core excited state $|C\rangle$, followed by the Auger decay to state $|A\rangle$ with decay rate $\Gamma_{Aug}$(see Fig. {\color{blue} 1}). Our studies show that with a bandwidth of the probe pulse approaching the energy spread of the wave packet, the Auger yields and spectra show quantum beats as a function of pump-probe time delay $\Delta t$. \\

Our studies are limited to "weak" probe pulses, in the sense that a perturbative treatment of the Auger process is applicable. The intensities of the applied probe field should be small enough, so that strong resonant coupling, inducing Rabi flopping between the resonantly coupled levels and resulting in the typical broadening and splitting of the Auger spectra {\color{blue}\cite{Rohringer2008,Demekhin2011a,Demekhin2011b,Rohringer2012,Demekhin2013a,muller2015}}. 
 can be neglected.  In other words, the intensities are chosen small enough, so that the Rabi period is much longer than the pulse duration of the probe pulse. \\

The outline of the paper is as follows: In the next section, the theoretical methods including the numerical method and the analytic solution for the model system are introduced, followed by section III that presents the numerical results and discussions; the possibility of extending this method to molecular system is discussed and numerical results on the CO molecule are given in section IV; section V gives a summary of this work.Atomic units (a.u.) are used throughout the paper, if not otherwise stated. 

\section{Theoretical Methods}  

The time-dependent wave packet propagation method for a few-level system {\color{blue} \cite{Pahl1996,Demekhin2011a,Demekhin2013a,Zhang2014}} is employed to evaluate the dynamics of the electronic states and the Auger electron spectrum. The total wave function $\Psi(t)$ can be expanded in
\begin{eqnarray}
\Psi(t)=a_{B_1}(t)|B_1\rangle+a_{B_2}(t)|B_2\rangle  \nonumber ~~~~~~~~~~~~~~~~~~~~~\\
+a_{C}(t)e^{-i\omega_0 t}|C\rangle
+\int a_{A}(\varepsilon,t)e^{-i\omega_0 t}|A,\varepsilon\rangle d\varepsilon,
\end{eqnarray}
where, $a_{B_1}(t)$, $a_{B_2}(t)$, $a_{C}(t))$ and $a_{A}(\varepsilon,t)$ are the time-dependent amplitudes of the levels $|B_1\rangle$,$|B_2\rangle$,$|C\rangle$ and $|A,\varepsilon\rangle$, respectively. In our study, we only treat the probe process explicitly and suppose that a coherent wave packet was prepared by an appropriate pump pulse (by for example Raman scattering) and at time $t=0$ the initial wave packet is $\Psi(t=0)=c_1|B_1\rangle+c_2e^{i\varphi_0}|B_2\rangle$. 
For the probe field, we suppose a weak linearly polarized electric field $G(t)=g_{0} g(t)cos(\omega_0 t)$ with pulse envelope $g(t)$, electric field strength $g_{0}$ and envelope peak at $t=\Delta t$. Inserting the total wave function into the time-dependent Schr\"odinger equation for the total Hamiltonian and implying the rotating wave approximation {\color{blue} \cite{Gamaly2011,Shore2011}} and the local approximation {\color{blue} \cite{Cederbaum1981,Domcke1991,Pahl1996,Demekhin2011b}} leads to the following equation determining the evolution of the expansion coefficients
\begin{eqnarray}
i\dot{\mathbf{\Psi}}_a(t)=\bar{\textbf{H}}(t)\mathbf{\Psi}_a(t),
\end{eqnarray}
where 
\begin{eqnarray}
\mathbf{\Psi}_a(t)=\mathbf{[}a_{B_1}(t),a_{B_2}(t),a_{C}(t),a_{A}(\varepsilon,t)\mathbf{]}^{T},
\end{eqnarray}
and
\begin{eqnarray}
\nonumber
\bar{\textbf{H}}(t) = ~~~~~~~~~~~~~~~~~~~~~~~~~~~~~~~~~~~~~~~~~~~~~~~~~~~~~~~~~~\\
\left( \begin{array}{cccc}
E_{B_1}   & 0           & D^\dag_1(t)                  &   0  \\
0         & E_{B_2}     & D^\dag_2(t)                  &   0  \\
D_1(t)    & D_2(t)      & E_C-\omega_0-i\frac{\Gamma_{Aug}}{2} &   0  \\
0         & 0           & V                            & E_A-\omega_0+\varepsilon
\end{array} \right).~~~~
\end{eqnarray}
The generally complex functions $D_{i}(t)=\Omega_ig(t)/2$ with the Rabi frequency $\Omega_i=g_0d_i$, $d_i$ is the transition dipole from state $|B_i\rangle$ to state $|C\rangle$ $(i=1,2)$; $V=\langle A,\varepsilon|1/\hat{r}_{12}|C \rangle$ is the Coulomb matrix element between the core excited state and the final ionic state; $\Gamma_{Aug}=2\pi|V|^2$ is the Auger decay width. Finally, the Auger electron spectrum with the residual ions in the ionic stat $|A\rangle$ is given by
\begin{equation}
\sigma_A(\varepsilon)=\lim _{t\rightarrow \infty} |a_A(\varepsilon,t)|^2.
\end{equation} 
The total Auger electron yield pertaining to the ionic channel $A$ can be computed as 
\begin{equation}
\sigma_A^T=\int {\sigma_A(\varepsilon)} d\varepsilon.
\end{equation} 
We suppose the pulse is short and weak and the direct ionization process is neglected in the theory. Therefore, only the Auger process contributes to the depletion of the bound states, so that the total Auger yield can also be computed by
\begin{equation}
\sigma_A^T=1-\lim _{t\rightarrow \infty}(|a_{B_1}(t)|^2 + |a_{B_2}(t)|^2).
\end{equation} 
The system of Eq.({\color{blue} 2}) was solved numerically employing Gaussian pulse $g(t)=e^{-2\ln2~t^{2}/\tau^{2}}$, where $\tau$ is the pulse duration at full width of half maximum of the field intensity. Since a perturbative probe pulse is employed, the system can also be approximately solved based on the time-dependent perturbation theory. The analytic solutions based on the second order time-dependent perturbation theory are given in detail in the Appendix. \\

A model system  is employed and we suppose $E_{CB_1}=210$ eV, $E_{CA}=200$ eV, $E_{B_2B_1}=0.2$ eV and $\Gamma_{Aug}=0.1$ eV (corresponding to a lifetime of 6.6 fs), that are typical in atomic and molecular soft x-ray induced excitations. The pulse duration $\tau$ is varied from several femtoseconds to tens of femtoseconds. In terms of the "weak" probe-pulse limit we suppose an Rabi frequencies of $\Omega_1=\Omega_2=0.0001~ a.u.$, corresponding to a Rabi period of about 1.5 ps, i.e. much longer than the considered pulse duration. For typical dipole matrix elements in the soft x-ray domain, this would correspond to a peak intensity in the range of 10$^{13}$-10$^{15}$ W/cm$^2$. The spectral shape below this intensity limit does not depend on the applied intensities, so that an integration over the spatial focus profile of the interaction region will not change the presented spectra.\\

We study different choices for the initial phase $\varphi_0$ and initial state populations $c_1^2$ and $c_2^2$. Since the energy gap $E_{B_2B_1}$ can be resolved from the quantum beating of the time delayed Auger spectra or from the non resonant Auger spectra for long pulses, we suppose this quantity is known a priori. Generally, the probe frequency $\omega_0$ can be chosen anywhere between $E_{CB_1}$ and $E_{CB_2}$. Choosing $\omega_0=(E_{CB_1}+E_{CB_2})/2=E_{CB_1}-E_{B_2B_1}/2=209.9$ eV allows for a more compact derivation of the explicit analytical expression for the reconstruction of the wave packet by perturbation theory and greatly simplifies the discussion of the results, but it is not a restriction on the applicability of the method. With this specific choice, the analytic expression of the total Auger yield is given by
\begin{eqnarray}
\nonumber
\sigma_A^T(\omega_0=E_{CB_1}-E_{B_2B_1}/2) ~~~~~~~~~~~~~~~~~~~~~~~~~~~~~~~~~~~~~~\\
\nonumber
\simeq\frac{\pi \tau^2}{8\ln2} e^{-\frac{\tau^2}{16\ln2} (E_{B_2B_1}^2-\Gamma_{Aug}^2)}[(c_1\Omega_1)^2 p(\tau)+(c_2\Omega_2)^2 p(\tau) \\
+2 c_1\Omega_1 c_2\Omega_2 q(\tau)\cos\varphi],~~~~~~~~~~~~~~~~~~~~~~~~~~~~~~~~~~~~~~~~~
\end{eqnarray} 
where $\varphi=mod(\varphi_0-E_{B_2B_1}\Delta t,2\pi)$ is the scaled phase, which includes the initial phase and the phase accumulated with time. We define the functions $q(\tau)=[Erf(\frac{\Gamma_{Aug}}{4\sqrt{\ln2}}\tau)-1]$ and $p(\tau)=Re\big\{e^{i\frac{E_{B_2B_1}\Gamma_{Aug}}{8\ln2}\tau^2}[Erf(i\frac{E_{B_2B_1}-i\Gamma_{Aug}}{4\sqrt{\ln2}}\tau)-1]\big\}$, where $Erf$ is the error function. Eq.(8) shows the exchange symmetry between $c_1\Omega_1$ and $c_2\Omega_2$. \\

Being derived from perturbation theory, the linear dependence of the scaled phase with respect to the time-delay is strictly valid only in the perturbative limit. In the strong coupling case, additional phase can be accumulated by Rabi flopping between the resonantly coupled levels and the expression for the scaled phase and Eq.(8) are no longer valid. By choosing the Rabi period of the probe pulse at least ten times longer than the pulse duration, we make sure to stay in the perturbative limit of the applied intensities. For the considered Rabi frequencies, our numerical calculations (valid also for higher probe-pulse intensities) and the analytic results obtained by perturbation theory are in good agreement. As can be directly seen, in the limit of long pulse duration, q($\tau$) tends towards zero and the Auger spectrum approaches the static limit of an incoherent sum of contributions from states B$_1$ and B$_2$. \\

Similarly, the analytic expression of the Auger electron spectrum is given by
\begin{eqnarray}
\nonumber
\sigma_A(\omega_0=E_{CB_1}-E_{B_2B_1}/2,\varepsilon) ~~~~~~~~~~~~~~~~~~~~~~~~~~~~~~~~~\\
\nonumber
\simeq \frac{\Gamma_{Aug}\tau^2}{16\ln2(\Delta^2+\frac{\Gamma_{Aug}^2}{4})}\times ~~~~~~~~~~~~~~~~~~~~~~~~~~~~~~~~~~~~~~~~\\
\nonumber
\big\{(c_1\Omega_1)^2e^{-\frac{\tau^2}{4\ln2}(\Delta+\frac{E_{B_2B_1}}{2})^2}+(c_2\Omega_2)^2e^{-\frac{\tau^2}{4\ln2}(\Delta-\frac{E_{B_2B_1}}{2})^2} \\
+2 c_1\Omega_1c_2\Omega_2\cos\varphi~e^{-\frac{\tau^2}{8\ln2}[(\Delta+\frac{E_{B_2B_1}}{2})^2+(\Delta-\frac{E_{B_2B_1}}{2})^2]}\big\},~~~~ 
\end{eqnarray}
where $\Delta=\varepsilon-E_{CA}$. There is a special case for Eq.(9) if $c_1\Omega_1=c_2\Omega_2$, then the spectrum will be symmetric with respect to $\Delta$. Eq.(9) can be further simplified if $\Delta=0$ or the spectrum is detected at the Auger energy 
$\varepsilon=E_{CA}=200$ eV, as
\begin{eqnarray}
\nonumber
\sigma_A(\omega_0=E_{CB_1}-E_{B_2B_1}/2,\varepsilon=E_{CA})= \frac{\tau^2}{4\ln2\Gamma_{Aug}}~~~~~~~\\
\times e^{-\frac{\tau^2}{16\ln2}E_{B_2B_1}^2}(c_1^2\Omega_1^2+c_2^2\Omega_2^2+2 c_1c_2\Omega_1\Omega_2\cos\varphi).~~~~~~~~~~
\end{eqnarray}

These analytic expressions allow to extract the initial state populations and phase. \\

\section{Numerical Results}

The total Auger yield and Auger electron spectra for the case with equal initial state populations $c_1^2=c_2^2=0.5$ are shown in Fig. {\color{blue}2}. Fig. {\color{blue}2a} shows the total Auger yield as a function of the pulse duration $\tau$ and the scaled phase $\varphi$, that relates to the time delay $\Delta t$. The total overall yield increases with longer pulse duration, since the peak Rabi frequency is kept constant as the pulse duration is increased, leading to larger pulse energies. As Fig. {\color{blue}2a} shows, the total Auger yield shows only a small dependence on $\varphi$ for long pulses. For long pulses durations with $\tau>25 fs$, the energy bandwidth ($\Delta\omega <$  0.1 eV )) of the applied probe pulse is small compared to the energy splitting $E_{B_2B_1}=0.2$ eV and the Auger lifetime of the upper state and the situation is approaching the static limit, i.e. no variation of the total yield on the time-delay (scaled phase) is observable. In the quasi-static case, the probe pulse has a well-defined frequency and is considerably detuned  by $\pm E_{B_2B_1}/2$ from the core-valence transition energies $E_{CB_1}$ and $E_{CB_2}$. The Auger spectral characteristics therefore are described by the static limit of the resonant Auger effect (nonradiative resonance scattering) {\color{blue} \cite{gel1999}}. The energy of the Auger electron follows a linear dispersion with respect of the incoming photon energy, i.e. two distinct peaks appear due to the  detuned by $\pm E_{B_2B_1}/2$ from the resonant limits of transitions $E_{CB_1}$ and $E_{CB_2}$. The two distinct peaks can be seen in the Auger-electron spectra shown in Figs.\ {\color{blue}2b} and {\color{blue}2c}, that shows the spectra as a function of pulse duration for scaled phases of $\varphi=0$ an $\varphi=\pi$, respectively.\\
 
The situation changes significantly when $\varphi$ varies from 0 to 2$\pi$ and the pulse duration $\tau$ is reduced below 13.5 fs: In that case, the bandwidth of the probe pulse approaches the energy splitting $E_{B_2B_1}$ between states $B_1$ and $B2$, and is large compared to the Auger life-time width. The broad bandwidth supports resonance scattering initiated from both states $B_1$ and $B2$ to the same final Auger-electron energy and interference of the two pathways starts to dominate the Auger-electron spectrum. The total Auger yield for the pulses with fixed pulse duration $\tau\sim 13.5 fs$ shows clear maxima and a minimum for scaled phases $\varphi=0$ or $2\pi$ and $\varphi=\pi$, respectively. This means that cleary constructive and destructive interference of the resonance scattering of the two initial states $B_1$ and $B2$ is observed. This result provides a way to extract the initial phase $\varphi_0$: Explicitly, the delay times $\Delta t_{max}$ or $\Delta t_{min}$ corresponding to the maximum or minimum of the total Auger yield for a fixed pulse duration $\tau<13.5$ fs have to be determined. Analyzing Eq.\ (8), the initial phase $\varphi_0$ can then be found by the condition $mod(\varphi_0-E_{B_2B_1}\Delta t_{max},2\pi)=0$ or $mod(\varphi_0-E_{B_2B_1}\Delta t_{min},2\pi)=\pi$. In principle, the initial state populations can also be extracted by fitting the total Auger yield for a given pulse duration by the expression given by Eq.(8). A simpler and more straightforward way to extract populations will be introduced in the following, analyzing the Auger-electron spectra.\\

Fig. {\color{blue}2b} and Fig. {\color{blue}2c} show the Auger electron spectra as a function of the pulse duration for the scaled phase $\varphi=0$ and $\varphi=\pi$, respectively, and highlight the emergence of the interference pattern as the pulse duration decreased. When the pulse is long, two isolated main peaks are observed around $\omega_0-E_{AB_1}=199.9$ eV and $\omega_0-E_{AB_2}=200.1$ eV, corresponding to the quasi-static resonant Auger spectra from the two states $|B_1\rangle$ and $|B_2\rangle$ with well-defined photon-energy detuning of $\pm E_{B_2B_1}$, respectively; with decreasing pulse duration, the two main peaks become broader and shift towards 200 eV. The two channels begin to interfere, and for $\tau\sim 13.5$ fs, the spectra are dominated by the interference pattern. As shown clearly in the plots, $\varphi=0$ and $\varphi=\pi$ correspond to the cases of constructive and destructive interference, respectively.\\

The Auger electron spectra as a function of the scaled phase $\varphi$ (related to the pump-probe time delay) are shown in Fig. {\color{blue}2d} for a probe-pulse duration of $\tau=13.5$ fs. In the main spectral region where $\varepsilon
\sim E_{CA}=200$ eV, the spectra show strong dependence on the scaled phase $\varphi$, maximum and minimum are observed around $\varphi=0$ (and $2\pi$) and $\pi$, respectively. Those results are in agreement with the perturbative predictions of Eq.\ (9). Bearing in mind that Eq.\ (9) reduces into the simpler expression of Eq.(10) at Auger-electron energies of $\varepsilon=E_{CA}=200$ eV, the initial wave packet can be easily reconstructed by examining the spectral Auger intensity at $\varepsilon=E_{CA}$ as a function of the delay time (phase $\varphi$): the initial phase $\varphi_0$ and the state populations can be extracted from the delaye times corresponding to the maximum or minimum of the spectral intensity and the modulation depth $\beta$ of the spectral intensity [defined as (max-min)/max], respectively. The same method for extracting the initial phase from the total Auger yield of Fig. {\color{blue}2a} can be applied to extract the initial phase $\varphi_0$ from the spectral intensityes at $\varepsilon=E_{CA}$. Analyzing Eq.\ (10), we find that the modulation depth $\beta$ satisfies the following equation

\begin{eqnarray}
\beta=\frac{4c_1d_1c_2d_2}{c_1^2d_1^2+c_2^2d_2^2+2c_1d_1c_2d_2},
\end{eqnarray} \\
which gives two solutions: $\frac{c_1d_1}{c_2d_2}=\frac{2-\beta-2\sqrt{1-\beta}}{\beta}$ and $\frac{c_2d_2}{c_1d_1}=\frac{2-\beta-2\sqrt{1-\beta}}{\beta}$. If the dipole ratio $\frac{d_1}{d_2}\neq 1$ and is known a priori, the initial state populations $c_1^2$ and $c_2^2$ can be uniquely determined with the additional constraint $c_1^2+c_2^2=1$.\\

The evolution of the modulation depth at the resonant Auger energy $\varepsilon=E_{CA}=200$ eV  is shown in Fig. {\color{blue}3} as a function of scaled phase for different initial state populations and for a pulse duration $\tau=13.5$ fs. The modulation depths for initial state occupations of $c_1^2=0.5$ ($c_2^2=0.5$), $c_1^2=0.9$ ($c_2^2=0.1$), $c_1^2=0.99$ ($c_2^2=0.01$) and $c_1^2=0.999$ ($c_2^2=0.001$) are $100.0\%$, $75.0\%$, $33.2\%$ and $11.9\%$, respectively. Even a low excitation with of $0.1\%$ from the ground state results in a significant modulation depth of $11.9\%$ in the delay-dependent Auger electron spectra, if the dipole transition strengths $d_1$ and $d_2$ to the intermediate core-excited state are comparable in size. Due to the exchange symmetry of Eqs.\ (8-10)  with respect to the product $c_1d_1$ and $c_2d_2$, the occupation probabilities $c_1^2$ and $c_2^2$ can not be unambiguously determined for $d_1=d_2$ in the considered case. By shifting the central frequency $\omega_0$ off the midpoint between transition energies $E_{CB_1}$ and $E_{B_2B_1}$, it is however possible to lift the ambiguity in the assignment of the occupation probabilities to states $B_1$ and $B_2$. A complete analysis of the delay-dependent Auger spectra then yields a unique reconstruction of the initial level occupancies and their relative phase.\\

The situation is different for initially occupied states that show different coupling strength to the intermediate core-excited state. Fig.\ {\color{blue}4} shows the Auger spectra  as a function of scaled phase (delay time) for a probe-pulse duration of $\tau=13.5$ fs for an initial state population of $c_1^2=0.9$ ($c_2^2=0.1$) for different ratios of the transition dipole moments $\frac{d_1}{d_2}$. The driving frequency of the probe field is still supposed midway between the core-excited resonances. 
Figs.\ {\color{blue}4a}, {\color{blue}4b}, {\color{blue}4c} and {\color{blue}4d} correspond to the cases with the dipole ratios of $\frac{d_1}{d_2}=\frac{\Omega_1}{\Omega_2}=\frac{1}{0.5}$, $\frac{1}{1}$, $\frac{1}{3}$ and $\frac{1}{4}$, respectively. 
As shown clearly, the Auger electron spectra in Fig. {\color{blue}4c} for $\frac{c_1d_1}{c_2d_2}= \frac{c_1\Omega_1}{c_2\Omega_2}=1$ (corresponding to $\frac{d_1}{d_2}=1/3$ and $c_1^2=0.9$, $c_2^2=0.1$) show a symmetry with respect to the energy$\varepsilon=E_{CA}=200$ eV, in consistence with Eq.\ (9). This symmetry is no longer prevalent in the other cases of Figs. {\color{blue}4a, b and d}), that present results for ratios $\frac{c_1d_1}{c_2d_2}\neq 1$ ($\frac{c_1d_1}{c_2d_2}$ of $\frac{3}{0.5}$, $\frac{3}{1}$ and $\frac{3}{4}$, respectively). 
Eq.\ (9), discussing the case of central probe frequency $\omega_0=E_{CB_1}-E_{B_2B_1}/2$, clearly shows that the Auger spectra do not show a symmetry with respect to the resonant Auger-energy $\varepsilon=E_{CA}$, if $\frac{c_1\Omega_1}{c_2\Omega_2}=\frac{c_1d_1}{c_2d_2} \neq 1$. This asymmetry of the Auger spectra allows to lift the unambiguity presented by the solutions of Eq. (11) for the modulation depth at $\varepsilon=E_{CA}$, that has two solutions for the ratios $\frac{c_1d_1}{c_2d_2}$.
If the partial Auger yield of energies $\varepsilon<E_{CA}$  ($\varepsilon>E_{CA}$) is dominant in the Auger-spectrum (the Auger-spectrum is shifted below (above) the resonance energy condition), then $\frac{c_1d_1}{c_2d_2}>1$ ($\frac{c_1d_1}{c_2d_2}<1$, which chooses one solution of Eq.\ (11). Therefore, the ratio $\frac{c_1d_1}{c_2d_2}$ can be uniquely determined. Moreover, if the ratio of the dipole transition strengths $\frac{d_1}{d_2}$ is known, the initial state occupancies $c_1^2$ and $c_2^2$ and their relative phase can be unambigyously recovered. Also in the more general case of the probe-pulse driving frequency $\omega_0 \neq E_{CB_1}-E_{B_2B_1}/2$, a similar reconstruction can be achieved. In that case, the delay-dependent modulation of the Auger spectral intensity has to be monitored at the energy
$\varepsilon=E_{CA}+\omega_0-\frac{E_{CB_1}+E_{CB_2}}{2}$ and similar analysis and reconstruction to that presented applies. \\  

\section{Application of quantum-beat Auger spectroscopy to molecules}  

Although the quantum-beat Auger spectroscopy was studied based on a relatively simple four-level model system, we note that such a four-level model system can be applied to a large class of real atomic or ionic systems. For example, Ar$^+$ would be a possible realistic system, with an energy splitting of the first two low-lying (spin-orbit) bound states of 0.18 eV and a 2$s$ excitation threshold of about 300 eV. The real power of the method will, undoubtedly, lie in the application to molecular systems, that can be prepared in a mixture of coherent electronic and vibrational wave packets. As in optical quantum-beat spectroscopy, a complete reconstruction of a complex wave packet, involving many different states and relative phases, is not realistically achievable. To discuss the challenges and complications involved, we present a numerical case study on quantum-beat Auger spectroscopy on the CO molecule. We suppose a relatively simple wave packet, consisting only of a mixture of the lowest two vibrational states in the electronic ground state of CO. Coupling of this electronic ground-state wave packet to the intermediate core-excited manifold by a femtosecond broadband x-ray pulse results in the excitation of a vibrational wave packet in the core-excited state. The electronic potential surfaces of ground- and core-excited state are not necessarily parallel, so that many different vibrational excited states in the core-excited manifold are involved in the resonance scattering process. The delay-dependent Auger spectrum will therefore also be sensitive to the intermediate core-excited valence-wave packets that are induced by resonant excitation from the electronic ground state. The probe process, with more than one intermediate level being involved, is therefore too complex to allow for a unique reconstruction of the initial ground-state vibrational wave packet. Nevertheless, quantum beat Auger spectroscopy can create some knowledge of the initial wave-packet. We discuss the case on numerical results obtained for the
CO molecule as an illustrative example.\\

We suppose that at time zero, CO is in a coherent superposition of vibrational states $|\nu=0\rangle$ and $|\nu=1\rangle$ (or $(|\nu=0\rangle+e^{i\phi_0}|\nu=1\rangle)/\sqrt{2}$ ) of the electronic ground state $X~ ^1\Sigma^+$. The energy difference of the first two vibrational states in state $X~ ^1\Sigma^+$ is $E_{\nu_{01}}$=0.266 eV. As in the model system treated in sections II and III, the probe pulse is triggered at time $\Delta t$ with respect to the pump pulse and has a driving frequency to resonantly couple to the intermediate core excited state C 1s$^{-1}\pi^*~^1\Pi$, roughly 287.4 eV above the electronic ground-state. The core-excited state has an Auger-decay width of $\Gamma$=0.08 eV and decays predominantly into the final ionic state $1\pi^{-1~2}\Sigma^+$ (see Fig. {\color{blue}5a}). We treat the system in the Born Oppenheimer approximation and restrict the electronic degrees of freedom to only these three states. The electronic potential curves are described by Morse potentials with parameters taken from Ref. {\color{blue} \cite{Skytt1997}}. The time-dependent wave packet propagation method for the molecular system is employed and details of this method for molecular systems can be found in Refs. {\color{blue} \cite{Pahl1996,Demekhin2011a}}. In our numerical simulation, we choose the probe-pulse frequency in the middle of the resonance frequencies ($\omega_0$=287.543 eV=287.41-0.266/2), the peak Rabi frequency of the probe pulse is supposed to be 0.0001 a.u. (corresponding to a Rabi period of about 1.5 ps). \\

Fig.\ {\color{blue}5b} shows the total Auger yield with respect to the pulse duration $\tau$ and the scaled phase $\varphi = mod(\varphi_0-E_{\nu_{01}}\Delta t,2\pi)$. For pulses $\tau<10$ fs, corresponding to the energy splitting of the vibrational states of the initial wave packet, the total Auger yield shows strong variation as a function of the pump-probe delay (scaled phase). The Auger yield shows little dependence on $\varphi$ for long pulses. These features are quite similar as that shown in Fig. {\color{blue}2a}. Fig. {\color{blue}5c} and Fig. {\color{blue}5d} show the vibrationally resolved Auger electron spectra for $\varphi=0$ and $\pi$, respectively, as a function of the pulse duration. As the plots show, for long pulses, the Auger spectra for both cases of $\varphi=0$ and $\varphi=\pi$  are quasi identical, because of little interference from the two initial states $|0\rangle$ and $|1\rangle$. The observed spectrum consists of many different resonant scattering channels, involving several vibrational levels of the core excited and final ionic electronic states. The spectra correspond to an 'incoherent' sum of resonance scattering from initial states $|\nu=0\rangle$ and $|\nu=1\rangle$. Decreasing the pulse duration, the pulses excite more vibrational states in the core excited manifold. The resonance scattering pathways initiating from vibrational states $|\nu=0\rangle$ and $|\nu=1\rangle$ of the ground electronic state interfere. Therefore, more complex wave packet dynamics is created, which is manifested in the energy shifts and splitting of the Auger spectra. The Auger spectra for the cases of $\varphi=0$ and $\varphi=\pi$ show significant differences for pulses of less than 10 fs, due to destructive (clearly visible for the case of $\varphi=0$ for energies above 271.1 eV) and constructive interference (clearly visible for the case of $\varphi=\pi$ for energies above 271.1eV) of the different scattering pathways. \\

The vibrationally resolved Auger electron spectra for pulses of $\tau$=10 fs (20 fs) as a function of the scaled phase are shown in Fig. {\color{blue}6b} (Fig. {\color{blue}6d}). To highlight the delay-dependent changes of the Auger spectra, we also show the  Auger spectra from initial vibrational states $|0\rangle$, $|1\rangle$ and their incoherent sum in Fig. {\color{blue}6a} and Fig. {\color{blue}6c}. As shown in Fig. {\color{blue}6d}, the Auger spectra show weak dependence on the pump-probe delay for $\tau$=20 fs. The spectra can be understood as the incoherent sum of spectra of initial states $|\nu=0\rangle$ and $|\nu=1\rangle$ shown in Fig. {\color{blue}6a}. The broad bandwidth of the probe pulse with 10 fs duration results in the excitation to many different vibrational levels in the core-excited electronic state and in a smearing of the energy resolution of the vibrational transitions, which results in the more complex structures shown in Fig. {\color{blue}6a} as compared to Fig. {\color{blue}6c}. 
Clearly strong interference effects of scattering pathways initiating from levels $|\nu=0\rangle$ and $|\nu=1\rangle$ are visible in Fig. 6b at Auger energies in the range of 270.8 to 271.2 eV, showing a large variation for the Auger spectrum as a function of pump-probe delay. Quantum beats can be clearly seen in the auger spectra. A reconstruction of the initial vibrational wave-packet, however, proofs as difficult, due to complex wave-packet dynamics excited in the core-excited state of the system. Quantum beat spectroscopy {\color{blue} \cite{hack1991,carter2000}} could be a potential technique to help us to clear the difficulties, further studies will be performed in the near future.\\

\section{Conclusions}  
We introduced quantum-beat Auger spectroscopy by discussing a relatively simple four-level model system. We suppose a coherent wave packet involving two low-lying states that was prepared by an appropriate pump pulse. This wave packet is subsequently probed by a weak, time-delayed ultrashort x-ray pulse that has near coupling to a core-excited state of the atomic or molecular system. The Auger spectra are then studied as a function of the duration of the probe pulse and the time delay. With a bandwidth of the probe pulse approaching the energy spread of the wave packet, the Auger yields and spectra show quantum beats as a function of pump-probe delay. An analytic theory for the quantum-beat Auger spectroscopy is also presented, which allows for the full reconstruction of the wave packet by analyzing the delay-dependent Auger spectra. The extension of this method to a more complex manifold of electronic and vibrational energy levels is possible, however, methods for complete reconstruction of the wave packet by quantum-beat Auger spectroscopy need to be further investigated. The techniques of creating controlled two-pulse emission at present-day FELs of different wavelength, pulse duration and relative time-delays are rapidly evolving {\color{blue} \cite{lutman2013,allaria2013,marinelli2015}}. Moreover, sub-fs precision measurements of the relative delay times of two-pulse FEL schemes {\color{blue} \cite{hartmann2014}} have been developed, so that the presented quantum beat pump-probe spectroscopy should be feasible at current FEL facilities.

\section{Appendix}  
Second order time dependent perturbation theory is employed to approximately solve the Hamiltonian of Eq.(4). In this case, the diagonal parts and the off diagonal parts of Eq.(4) are considered as unperturbed and perturbed Hamiltonians, respectively. The expansion coefficients $a_n(t)$ ($n$=$B_1$, $B_2$ or $A$) can be expanded in powers of the interactions (second order) {\color{blue} \cite{shankar2012}},
\begin{equation}
a_n(t)\simeq a_n^{(0)}+a_n^{(1)}(t)+a_n^{(2)}(t),
\end{equation}
where $a_n^{(m)}$ is the $m^{th}$ order amplitude of state $n$. An appropriate pump pulse creates the two state wave packet with relative phase $\varphi_0$, the wave packet evolves freely as $\Psi(\Delta t)=c_1|B_1\rangle+c_2e^{i\varphi}|B_2\rangle$, where the relative phase $\varphi=\varphi_0-E_{B_2B_1}\Delta t$ and $\Delta t$ is defined as the time delay from the pump pulse envelope center. Supposing the weak probe pulse is applied at time $\Delta t$, the perturbation expressions of $a_n^{(m)}$ {\color{blue} \cite{shankar2012}} can be employed, and we can easily get the zeroth and first order amplitudes: $a_{B_1}^{(0)}=c_1$, $a_{B_2}^{(0)}=c_2e^{i\varphi}$, $a_{B_1}^{(1)}=a_{B_2}^{(1)}=a_{A}^{(0)}=a_{A}^{(1)}=0$. Let us define  $\omega_{AC}=-\omega_{CA}=E_{AC}+\varepsilon-i\frac{\Gamma_{Aug}}{2}$, $\omega_{CB_i}=-\omega_{B_iC}=E_{CB_i}-\omega_0-i\frac{\Gamma_{Aug}}{2}$, and $\omega_{AB_i}=-\omega_{B_iA}=E_{AB_i}-\omega_0+\varepsilon$ $(i=1,2)$, the second order amplitudes,
\begin{eqnarray}
\nonumber
a_{B_1}^{(2)}(t\to\infty) ~~~~~~~~~~~~~~~~~~~~~~~~~~~~~~~~~~~~~~~~~~~~~~~~~~~~~~~~~\\
\nonumber
=-c_1\int_{-\infty}^{t}dt'e^{i\omega_{B_1C}t'}D_1(t')\int_{-\infty}^{t'}dt''e^{i\omega_{CB_1}t''}D_1(t'') ~~~~\\
\nonumber
-c_2e^{i\varphi}\int_{-\infty}^{t}dt'e^{i\omega_{B_1C}t'}D_1(t')\int_{-\infty}^{t'}dt''e^{i\omega_{CB_2}t''}D_2(t'') ~~~\\
\nonumber
=-\frac{c_1\Omega_1^2\tau}{8\sqrt{2\pi\ln2}}\int_{-\infty}^{t}dt'e^{i\omega_{B_1C}t'}g(t')\int_{-\infty}^{t'}dt''e^{i\omega_{CB_1}t''} ~~~~\\
\nonumber
\times\int_{-\infty}^{\infty}d\omega e^{-\frac{\tau^2\omega^2}{8\ln2}}e^{i\omega t''} ~~~~~~~~~~~~~~~~~~~~~~~~~~~~~~~~~~~~~~~~~~~\\
\nonumber
-\frac{c_2e^{i\varphi}\Omega_1\Omega_2\tau}{8\sqrt{2\pi\ln2}}\int_{-\infty}^{t}dt'e^{i\omega_{B_1C}t'}g(t')\int_{-\infty}^{t'}dt''e^{i\omega_{CB_2}t''} ~~~~\\
\nonumber
\times\int_{-\infty}^{\infty}d\omega e^{-\frac{\tau^2\omega^2}{8\ln2}}e^{i\omega t''}  ~~~~~~~~~~~~~~~~~~~~~~~~~~~~~~~~~~~~~~~~~~\\
\nonumber
=i\frac{c_1\Omega_1^2\tau}{8\sqrt{2\pi\ln2}}\int_{-\infty}^{\infty}d\omega e^{-\frac{\tau^2\omega^2}{8\ln2}}\int_{-\infty}^{t}dt'\frac{e^{i\omega t'}g(t')}{\omega_{CB_1}+\omega} ~~~~~~~~~~~\\
\nonumber
+i\frac{c_2e^{i\varphi}\Omega_1\Omega_2\tau}{8\sqrt{2\pi\ln2}}\int_{-\infty}^{\infty}d\omega e^{-\frac{\tau^2\omega^2}{8\ln2}}\int_{-\infty}^{t}dt'\frac{e^{i(E_{B_1B_2}+\omega)t'}g(t')}{\omega_{CB_2}+\omega} \\
\nonumber
=i\frac{c_1\Omega_1^2\tau}{16\ln2}\int_{-\infty}^{\infty}d\omega\frac{e^{-\frac{\tau^2\omega^2}{8\ln2}}}{\omega_{CB_1}+\omega}\tau e^{-\frac{\tau^2\omega^2}{8\ln2}} ~~~~~~~~~~~~~~~~~~~\\
\nonumber
+i\frac{c_2e^{i\varphi}\Omega_1\Omega_2\tau}{16\ln2}\int_{-\infty}^{\infty}d\omega\frac{e^{-\frac{\tau^2\omega^2}{8\ln2}}}{\omega_{CB_2}+\omega}\tau e^{-\frac{\tau^2(\omega-E_{B_2B_1})^2}{8\ln2}} ~~~\\
\nonumber
=\frac{c_1\pi}{16\ln2}\tau^2\Omega_1^2e^{-\frac{\tau^2\omega_{CB_1}^2}{4\ln2}}[Erf(i\frac{\tau}{2\sqrt{\ln2}}\omega_{CB_1})-1] ~~~~~~~~\\
\nonumber
+ \frac{c_2e^{i\varphi}\pi}{16\ln2}\tau^2\Omega_1\Omega_2e^{-\frac{1}{16\ln2}\tau^2E_{B_2B_1}^2}e^{-\frac{\tau^2}{4\ln2}(\frac{\omega_{CB_1}+\omega_{CB_2}}{2})^2} ~~~~\\
\times [Erf(i\frac{\tau}{2\sqrt{\ln2}}\frac{\omega_{CB_1}+\omega_{CB_2}}{2})-1], ~~~~~~~~~~~~~~~~~~~~~~~~
\end{eqnarray}
where Erf is the error function. Note that $t'$ and $t''$ in the integration are relative to the probe pulse envelope center. The two parts included in the above expression correspond to the second order processes $|B_1\rangle\to|C\rangle\to|B_1\rangle$ and $|B_2\rangle\to|C\rangle\to|B_1\rangle$, respectively. Similarly
\begin{eqnarray}
\nonumber
a_{B_2}^{(2)}(t\to\infty) ~~~~~~~~~~~~~~~~~~~~~~~~~~~~~~~~~~~~~~~~~~~~~~~~~~~~~~~~~\\
\nonumber
=-c_2e^{i\varphi}\int_{-\infty}^{t}dt'e^{i\omega_{B_2C}t'}D_2(t')\int_{-\infty}^{t'}dt''e^{i\omega_{CB_2}t''}D_2(t'') \\
\nonumber
-c_1\int_{-\infty}^{t}dt'e^{i\omega_{B_2C}t'}D_2(t')\int_{-\infty}^{t'}dt''e^{i\omega_{CB_1}t''}D_1(t'') ~~~~~~\\
\nonumber
=\frac{c_2e^{i\varphi}\pi}{16\ln2}\tau^2\Omega_2^2e^{-\frac{\tau^2\omega_{CB_2}^2}{4\ln2}}[Erf(i\frac{\tau}{2\sqrt{\ln2}}\omega_{CB_2})-1] ~~~~~~~~\\
\nonumber
+ \frac{c_1\pi}{16\ln2}\tau^2\Omega_1\Omega_2e^{-\frac{1}{16\ln2}\tau^2E_{B_2B_1}^2}e^{-\frac{\tau^2}{4\ln2}(\frac{\omega_{CB_1}+\omega_{CB_2}}{2})^2} ~~~~\\
\times [Erf(i\frac{\tau}{2\sqrt{\ln2}}\frac{\omega_{CB_1}+\omega_{CB_2}}{2})-1], ~~~~~~~~~~~~~~~~~~~~~~~
\end{eqnarray}
and
\begin{eqnarray}
\nonumber
a_A^{(2)}(t\to\infty) ~~~~~~~~~~~~~~~~~~~~~~~~~~~~~~~~~~~~~~~~~~~~~~~~~~~~~~~~~\\
\nonumber
=-c_1\int_{-\infty}^{t}dt'e^{i\omega_{AC}t'}V\int_{-\infty}^{t'}dt''e^{i\omega_{CB_2}t''}D_1(t'') ~~~~~~~~~~~\\
\nonumber
-c_2e^{i\varphi}\int_{-\infty}^{t}dt'e^{i\omega_{AC}t'}V\int_{-\infty}^{t'}dt''e^{i\omega_{CB_2}t''}D_2(t'') ~~~~~~~~~~\\
\nonumber
=-\frac{c_1\tau\Omega_1 V}{4\sqrt{2\pi\ln2}}\int_{-\infty}^{t}dt'e^{i\omega_{AC}t'}\int_{-\infty}^{t'}dt''e^{i\omega_{CB_1}t''} ~~~~~~~~~~~~~\\
\nonumber
\times\int_{-\infty}^{\infty}d\omega e^{-\frac{\tau^2\omega^2}{8\ln2}}e^{i\omega t''}  ~~~~~~~~~~~~~~~~~~~~~~~~~~~~~~~~~~~~~~~~~~~~\\
\nonumber
-\frac{c_2e^{i\varphi}\tau\Omega_2 V}{4\sqrt{2\pi\ln2}}\int_{-\infty}^{t}dt'e^{i\omega_{AC}t'}\int_{-\infty}^{t'}dt''e^{i\omega_{CB_2}t''} ~~~~~~~~~~~~~\\
\nonumber
\times\int_{-\infty}^{\infty}d\omega e^{-\frac{\tau^2\omega^2}{8\ln2}}e^{i\omega t''}  ~~~~~~~~~~~~~~~~~~~~~~~~~~~~~~~~~~~~~~~~~~~~\\
\nonumber
=i\frac{c_1\tau\Omega_1V}{4\sqrt{2\pi\ln2}}\int_{-\infty}^{\infty}d\omega e^{-\frac{\tau^2\omega^2}{8\ln2}}\int_{-\infty}^{t}dt'\frac{e^{i(\omega_{AB_1}+\omega)t'}}{\omega_{CB_1}+\omega} ~~~~~~~~~\\
\nonumber
+i\frac{c_2e^{i\varphi}\tau\Omega_2V}{4\sqrt{2\pi\ln2}}\int_{-\infty}^{\infty}d\omega e^{-\frac{\tau^2\omega^2}{8\ln2}}\int_{-\infty}^{t}dt'\frac{e^{i(\omega_{AB_2}+\omega)t'}}{\omega_{CB_2}+\omega} ~~~~~~~~\\
\nonumber
=i\frac{c_1\tau\Omega_1V_1}{4\sqrt{2\pi\ln2}}\int_{-\infty}^{\infty}d\omega\frac{e^{-\frac{\tau^2\omega^2}{8\ln2}}}{\omega_{CB_1}+\omega}2\pi\delta(\omega+\omega_{AB_1},0) ~~~~~~~~~\\
\nonumber
+i\frac{c_2e^{i\varphi}\tau\Omega_2V}{4\sqrt{2\pi\ln2}}\int_{-\infty}^{\infty}d\omega \frac{e^{-\frac{\tau^2\omega^2}{8\ln2}}}{\omega_{CB_2}+\omega}2\pi\delta(\omega+\omega_{AB_2},0) ~~~~~~~~\\
\nonumber
=-i\frac{c_1\tau\sqrt{\pi}\Omega_1V}{2\sqrt{2\ln2}}\frac{e^{-\frac{\tau^2\omega_{AB_1}^2}{8\ln2}}}{\omega_{AC}} ~~~~~~~~~~~~~~~~~~~~~~~~~~~~~~~~~~~~~\\
-i\frac{c_2e^{i\varphi}\tau\sqrt{\pi}\Omega_2V}{2\sqrt{2\ln2}}\frac{e^{-\frac{\tau^2\omega_{AB_2}^2}{8\ln2}}}{\omega_{AC}}, ~~~~~~~~~~~~~~~~~~~~~~~~~~~~~~~~~~~
\end{eqnarray}
whose two parts are contributed by state $|B_1\rangle$ and $|B_2\rangle$, respectively. Note that the integral expression of the Gaussian function in the frequency domain helps us to separate the variables and makes the analytic integration possible. Finally, the the total electron depletion (equals to the total Auger yield in the present case) can be calculated as 
\begin{eqnarray}
\nonumber
\sigma_A^T
\simeq 1-\lim _{t\rightarrow \infty}(|a_{B_1}^{(0)}(t)+a_{B_1}^{(2)}(t)|^2 + |a_{B_2}^{(0)}(t)+a_{B_2}^{(2)}(t)|^2 ) ~~~~~~\\
\nonumber
\simeq -\lim _{t\rightarrow \infty}(2Re(a_{B_1}^{(0)*}(t)a_{B_1}^{(2)}(t)+2Re(a_{B_2}^{(0)*}(t)a_{B_2}^{(2)}(t))  ~~~~~~~~~~\\
\nonumber
=\frac{\pi \tau^2}{8\ln2}(c_1\Omega_1)^2 Re(e^{-\frac{\tau^2}{4\ln2} \omega_{CB_1}^2}[Erf(i\frac{\tau}{2\sqrt{\ln2}}\omega_{CB_1})-1])~~~~\\
\nonumber
+\frac{\pi \tau^2}{8\ln2}(c_2\Omega_2)^2 Re(e^{-\frac{\tau^2}{4\ln2} \omega_{CB_2}^2}[Erf(i\frac{\tau}{2\sqrt{\ln2}}\omega_{CB_2})-1]) ~~~~\\
\nonumber
+\frac{\pi \tau^2}{4\ln2}c_1\Omega_1c_2\Omega_2\cos\varphi e^{-\frac{\tau^2}{16\ln2} E_{B_2B_1}^2}Re(e^{-\frac{\tau^2}{4\ln2}(\frac{\omega_{CB_1}+\omega_{CB_2}}{2})^2} \\
\times[Erf(i\frac{\tau}{2\sqrt{\ln2}}\frac{\omega_{CB_1}+\omega_{CB_2}}{2})-1]). ~~~~~~~~~~~~~~~~~~~~~~~~~~~~
\end{eqnarray} 
The three parts in the above expression have very clear correspondences, to the bound states $|B_1\rangle$, $|B_2\rangle$ and their coherence, respectively. The interference term [last term in Eq.(16)] modulated with the relative phase $\varphi$, that is determined by the time delay. Note that Erf(z$^*$)=Erf(z)$^*$, when $\omega_0=E_{CB_1}-E_{B_2B_1}/2$, we have $i\omega_{CB_1}=(i\omega_{CB_2})^*$, so $e^{-\frac{\tau^2}{4\ln2} \omega_{CB_1}^2}=(e^{-\frac{\tau^2}{4\ln2} \omega_{CB_2}^2})^*$ and $Erf(i\frac{\tau}{2\sqrt{\ln2}}\omega_{CB_1})=Erf(i\frac{\tau}{2\sqrt{\ln2}}\omega_{CB_2})^*$. Due to these symmetry relations, a simplified expression for the case $\omega_0=E_{CB_1}-E_{B_2B_1}/2$ can be derived (see Eq. (8) of main text). The Auger electron spectrum can be calculated as
\begin{eqnarray}
\nonumber
\sigma_A(\varepsilon)
\simeq \lim _{t\rightarrow \infty} |a_A^{(2)}(\varepsilon,t)|^2~~~~~~~~~~~~~~~~~~~~~~~~~~~~~~~~~~~~~~~~~~~ \\
\nonumber
=\frac{\Gamma_{Aug}\tau^2}{16\ln2|\omega_{AC}|^2}[(c_1\Omega_1)^2e^{-\frac{\tau^2}{4\ln2}\omega_{AB_1}^2}+(c_2\Omega_2)^2e^{-\frac{\tau^2}{4\ln2}\omega_{AB_2}^2} \\
+2\cos\varphi c_1\Omega_1c_2\Omega_2e^{-\frac{\tau^2}{8\ln2}(\omega_{AB_1}^2+\omega_{AB_2}^2)}], ~~~~~~~~~~~~~~~~~~~~~~~ 
\end{eqnarray}
whose three parts also possess the very clear interpretations, and are the same as that of Eq.(16).

\bibliographystyle{apsrev4-1}
\bibliography{refbib}

\begin{thebibliography}{48}%
\makeatletter
\providecommand \@ifxundefined [1]{%
 \@ifx{#1\undefined}
}%
\providecommand \@ifnum [1]{%
 \ifnum #1\expandafter \@firstoftwo
 \else \expandafter \@secondoftwo
 \fi
}%
\providecommand \@ifx [1]{%
 \ifx #1\expandafter \@firstoftwo
 \else \expandafter \@secondoftwo
 \fi
}%
\providecommand \natexlab [1]{#1}%
\providecommand \enquote  [1]{``#1''}%
\providecommand \bibnamefont  [1]{#1}%
\providecommand \bibfnamefont [1]{#1}%
\providecommand \citenamefont [1]{#1}%
\providecommand \href@noop [0]{\@secondoftwo}%
\providecommand \href [0]{\begingroup \@sanitize@url \@href}%
\providecommand \@href[1]{\@@startlink{#1}\@@href}%
\providecommand \@@href[1]{\endgroup#1\@@endlink}%
\providecommand \@sanitize@url [0]{\catcode `\\12\catcode `\$12\catcode
  `\&12\catcode `\#12\catcode `\^12\catcode `\_12\catcode `\%12\relax}%
\providecommand \@@startlink[1]{}%
\providecommand \@@endlink[0]{}%
\providecommand \url  [0]{\begingroup\@sanitize@url \@url }%
\providecommand \@url [1]{\endgroup\@href {#1}{\urlprefix }}%
\providecommand \urlprefix  [0]{URL }%
\providecommand \Eprint [0]{\href }%
\providecommand \doibase [0]{http://dx.doi.org/}%
\providecommand \selectlanguage [0]{\@gobble}%
\providecommand \bibinfo  [0]{\@secondoftwo}%
\providecommand \bibfield  [0]{\@secondoftwo}%
\providecommand \translation [1]{[#1]}%
\providecommand \BibitemOpen [0]{}%
\providecommand \bibitemStop [0]{}%
\providecommand \bibitemNoStop [0]{.\EOS\space}%
\providecommand \EOS [0]{\spacefactor3000\relax}%
\providecommand \BibitemShut  [1]{\csname bibitem#1\endcsname}%
\let\auto@bib@innerbib\@empty
\bibitem [{\citenamefont {Mehlhorn}(1998)}]{Mehlhorn1998}%
  \BibitemOpen
  \bibfield  {author} {\bibinfo {author} {\bibfnamefont {W.}~\bibnamefont
  {Mehlhorn}},\ }\href@noop {} {\bibfield  {journal} {\bibinfo  {journal} {J.
  Electron Spectrosc. Relat. Phenom.}\ }\textbf {\bibinfo {volume} {93}},\
  \bibinfo {pages} {1} (\bibinfo {year} {1998})}\BibitemShut {NoStop}%
\bibitem [{\citenamefont {Chang}(1971)}]{chang1971}%
  \BibitemOpen
  \bibfield  {author} {\bibinfo {author} {\bibfnamefont {C.~C.}\ \bibnamefont
  {Chang}},\ }\href@noop {} {\bibfield  {journal} {\bibinfo  {journal} {Surf.
  Sci.}\ }\textbf {\bibinfo {volume} {25}},\ \bibinfo {pages} {53} (\bibinfo
  {year} {1971})}\BibitemShut {NoStop}%
\bibitem [{\citenamefont {Weissmann}\ and\ \citenamefont
  {M{\"u}ller}(1981)}]{weissmann1981}%
  \BibitemOpen
  \bibfield  {author} {\bibinfo {author} {\bibfnamefont {R.}~\bibnamefont
  {Weissmann}}\ and\ \bibinfo {author} {\bibfnamefont {K.}~\bibnamefont
  {M{\"u}ller}},\ }\href@noop {} {\bibfield  {journal} {\bibinfo  {journal}
  {Surf. Sci. Rep.}\ }\textbf {\bibinfo {volume} {1}},\ \bibinfo {pages} {251}
  (\bibinfo {year} {1981})}\BibitemShut {NoStop}%
\bibitem [{\citenamefont {Moretti}(1998)}]{moretti1998}%
  \BibitemOpen
  \bibfield  {author} {\bibinfo {author} {\bibfnamefont {G.}~\bibnamefont
  {Moretti}},\ }\href@noop {} {\bibfield  {journal} {\bibinfo  {journal} {J.
  Electron. Spectrosc. Relat. Phenom.}\ }\textbf {\bibinfo {volume} {95}},\
  \bibinfo {pages} {95} (\bibinfo {year} {1998})}\BibitemShut {NoStop}%
\bibitem [{\citenamefont {Carlson}(2013)}]{carlson2013}%
  \BibitemOpen
  \bibfield  {author} {\bibinfo {author} {\bibfnamefont {T.}~\bibnamefont
  {Carlson}},\ }\href@noop {} {\emph {\bibinfo {title} {Photoelectron and Auger
  spectroscopy}}}\ (\bibinfo  {publisher} {Springer Science \& Business
  Media},\ \bibinfo {year} {2013})\BibitemShut {NoStop}%
\bibitem [{\citenamefont {Ullrich}\ \emph {et~al.}(2012)\citenamefont
  {Ullrich}, \citenamefont {Rudenko},\ and\ \citenamefont
  {Moshammer}}]{Ullrich2012}%
  \BibitemOpen
  \bibfield  {author} {\bibinfo {author} {\bibfnamefont {J.}~\bibnamefont
  {Ullrich}}, \bibinfo {author} {\bibfnamefont {A.}~\bibnamefont {Rudenko}}, \
  and\ \bibinfo {author} {\bibfnamefont {R.}~\bibnamefont {Moshammer}},\
  }\href@noop {} {\bibfield  {journal} {\bibinfo  {journal} {Annu. Rev. Phys.
  Chem.}\ }\textbf {\bibinfo {volume} {63}},\ \bibinfo {pages} {635} (\bibinfo
  {year} {2012})}\BibitemShut {NoStop}%
\bibitem [{\citenamefont {Rohringer}\ and\ \citenamefont
  {Santra}(2008)}]{Rohringer2008}%
  \BibitemOpen
  \bibfield  {author} {\bibinfo {author} {\bibfnamefont {N.}~\bibnamefont
  {Rohringer}}\ and\ \bibinfo {author} {\bibfnamefont {R.}~\bibnamefont
  {Santra}},\ }\href@noop {} {\bibfield  {journal} {\bibinfo  {journal} {Phys.
  Rev. A}\ }\textbf {\bibinfo {volume} {77}},\ \bibinfo {pages} {053404}
  (\bibinfo {year} {2008})}\BibitemShut {NoStop}%
\bibitem [{\citenamefont {Liu}\ \emph {et~al.}(2010)\citenamefont {Liu},
  \citenamefont {Sun}, \citenamefont {Wang} \emph {et~al.}}]{liu2010}%
  \BibitemOpen
  \bibfield  {author} {\bibinfo {author} {\bibfnamefont {J.-C.}\ \bibnamefont
  {Liu}}, \bibinfo {author} {\bibfnamefont {Y.-P.}\ \bibnamefont {Sun}},
  \bibinfo {author} {\bibfnamefont {C.-K.}\ \bibnamefont {Wang}},  \emph
  {et~al.},\ }\href@noop {} {\bibfield  {journal} {\bibinfo  {journal} {Phys.
  Rev. A}\ }\textbf {\bibinfo {volume} {81}},\ \bibinfo {pages} {043412}
  (\bibinfo {year} {2010})}\BibitemShut {NoStop}%
\bibitem [{\citenamefont {Kanter}\ \emph {et~al.}(2011)\citenamefont {Kanter},
  \citenamefont {Krassig}, \citenamefont {Li} \emph {et~al.}}]{Kanter2011}%
  \BibitemOpen
  \bibfield  {author} {\bibinfo {author} {\bibfnamefont {E.~P.}\ \bibnamefont
  {Kanter}}, \bibinfo {author} {\bibfnamefont {B.}~\bibnamefont {Krassig}},
  \bibinfo {author} {\bibfnamefont {Y.}~\bibnamefont {Li}},  \emph {et~al.},\
  }\href@noop {} {\bibfield  {journal} {\bibinfo  {journal} {Phys. Rev. Lett.}\
  }\textbf {\bibinfo {volume} {107}},\ \bibinfo {pages} {233001} (\bibinfo
  {year} {2011})}\BibitemShut {NoStop}%
\bibitem [{\citenamefont {Cederbaum}\ \emph {et~al.}(2011)\citenamefont
  {Cederbaum}, \citenamefont {Chiang}, \citenamefont {Demekhin} \emph
  {et~al.}}]{Cederbaum2011a}%
  \BibitemOpen
  \bibfield  {author} {\bibinfo {author} {\bibfnamefont {L.~S.}\ \bibnamefont
  {Cederbaum}}, \bibinfo {author} {\bibfnamefont {Y.-C.}\ \bibnamefont
  {Chiang}}, \bibinfo {author} {\bibfnamefont {P.~V.}\ \bibnamefont
  {Demekhin}},  \emph {et~al.},\ }\href@noop {} {\bibfield  {journal} {\bibinfo
   {journal} {Phys. Rev. Lett.}\ }\textbf {\bibinfo {volume} {106}},\ \bibinfo
  {pages} {123001} (\bibinfo {year} {2011})}\BibitemShut {NoStop}%
\bibitem [{\citenamefont {Demekhin}\ \emph {et~al.}(2011)\citenamefont
  {Demekhin}, \citenamefont {Chiang},\ and\ \citenamefont
  {Cederbaum}}]{Demekhin2011a}%
  \BibitemOpen
  \bibfield  {author} {\bibinfo {author} {\bibfnamefont {P.~V.}\ \bibnamefont
  {Demekhin}}, \bibinfo {author} {\bibfnamefont {Y.-C.}\ \bibnamefont
  {Chiang}}, \ and\ \bibinfo {author} {\bibfnamefont {L.~S.}\ \bibnamefont
  {Cederbaum}},\ }\href@noop {} {\bibfield  {journal} {\bibinfo  {journal}
  {Phys. Rev. A}\ }\textbf {\bibinfo {volume} {84}},\ \bibinfo {pages} {033417}
  (\bibinfo {year} {2011})}\BibitemShut {NoStop}%
\bibitem [{\citenamefont {Demekhin}\ and\ \citenamefont
  {Cederbaum}(2011)}]{Demekhin2011b}%
  \BibitemOpen
  \bibfield  {author} {\bibinfo {author} {\bibfnamefont {P.~V.}\ \bibnamefont
  {Demekhin}}\ and\ \bibinfo {author} {\bibfnamefont {L.~S.}\ \bibnamefont
  {Cederbaum}},\ }\href@noop {} {\bibfield  {journal} {\bibinfo  {journal}
  {Phys. Rev. A}\ }\textbf {\bibinfo {volume} {83}},\ \bibinfo {pages} {023422}
  (\bibinfo {year} {2011})}\BibitemShut {NoStop}%
\bibitem [{\citenamefont {Rohringer}\ and\ \citenamefont
  {Santra}(2012)}]{Rohringer2012}%
  \BibitemOpen
  \bibfield  {author} {\bibinfo {author} {\bibfnamefont {N.}~\bibnamefont
  {Rohringer}}\ and\ \bibinfo {author} {\bibfnamefont {R.}~\bibnamefont
  {Santra}},\ }\href@noop {} {\bibfield  {journal} {\bibinfo  {journal} {Phys.
  Rev. A}\ }\textbf {\bibinfo {volume} {86}},\ \bibinfo {pages} {043434}
  (\bibinfo {year} {2012})}\BibitemShut {NoStop}%
\bibitem [{\citenamefont {Demekhin}\ and\ \citenamefont
  {Cederbaum}(2013)}]{Demekhin2013a}%
  \BibitemOpen
  \bibfield  {author} {\bibinfo {author} {\bibfnamefont {P.~V.}\ \bibnamefont
  {Demekhin}}\ and\ \bibinfo {author} {\bibfnamefont {L.~S.}\ \bibnamefont
  {Cederbaum}},\ }\href@noop {} {\bibfield  {journal} {\bibinfo  {journal} {J.
  Phys. B}\ }\textbf {\bibinfo {volume} {46}},\ \bibinfo {pages} {164008}
  (\bibinfo {year} {2013})}\BibitemShut {NoStop}%
\bibitem [{\citenamefont {Muller}\ and\ \citenamefont
  {Demekhin}(2015)}]{muller2015}%
  \BibitemOpen
  \bibfield  {author} {\bibinfo {author} {\bibfnamefont {A.~D.}\ \bibnamefont
  {Muller}}\ and\ \bibinfo {author} {\bibfnamefont {P.~V.}\ \bibnamefont
  {Demekhin}},\ }\href@noop {} {\bibfield  {journal} {\bibinfo  {journal} {J.
  Phys. B}\ }\textbf {\bibinfo {volume} {48}},\ \bibinfo {pages} {075602}
  (\bibinfo {year} {2015})}\BibitemShut {NoStop}%
\bibitem [{\citenamefont {Chatterjee}\ and\ \citenamefont
  {Nakajima}(2015)}]{Chatterjee2015}%
  \BibitemOpen
  \bibfield  {author} {\bibinfo {author} {\bibfnamefont {S.}~\bibnamefont
  {Chatterjee}}\ and\ \bibinfo {author} {\bibfnamefont {T.}~\bibnamefont
  {Nakajima}},\ }\href@noop {} {\bibfield  {journal} {\bibinfo  {journal}
  {Phys. Rev. A}\ }\textbf {\bibinfo {volume} {91}},\ \bibinfo {pages} {043413}
  (\bibinfo {year} {2015})}\BibitemShut {NoStop}%
\bibitem [{\citenamefont {McFarland}\ \emph {et~al.}(2014)\citenamefont
  {McFarland}, \citenamefont {Farrell}, \citenamefont {Miyabe} \emph
  {et~al.}}]{McFarland2014}%
  \BibitemOpen
  \bibfield  {author} {\bibinfo {author} {\bibfnamefont {B.~K.}\ \bibnamefont
  {McFarland}}, \bibinfo {author} {\bibfnamefont {J.~P.}\ \bibnamefont
  {Farrell}}, \bibinfo {author} {\bibfnamefont {S.}~\bibnamefont {Miyabe}},
  \emph {et~al.},\ }\href@noop {} {\bibfield  {journal} {\bibinfo  {journal}
  {Nat Commun}\ }\textbf {\bibinfo {volume} {5}},\ \bibinfo {pages} {4235}
  (\bibinfo {year} {2014})}\BibitemShut {NoStop}%
\bibitem [{\citenamefont {Yudin}\ \emph {et~al.}(2005)\citenamefont {Yudin},
  \citenamefont {Chelkowski}, \citenamefont {Itatani} \emph
  {et~al.}}]{Yudin2005}%
  \BibitemOpen
  \bibfield  {author} {\bibinfo {author} {\bibfnamefont {G.~L.}\ \bibnamefont
  {Yudin}}, \bibinfo {author} {\bibfnamefont {S.}~\bibnamefont {Chelkowski}},
  \bibinfo {author} {\bibfnamefont {J.}~\bibnamefont {Itatani}},  \emph
  {et~al.},\ }\href@noop {} {\bibfield  {journal} {\bibinfo  {journal} {Phys.
  Rev. A}\ }\textbf {\bibinfo {volume} {72}},\ \bibinfo {pages} {051401}
  (\bibinfo {year} {2005})}\BibitemShut {NoStop}%
\bibitem [{\citenamefont {Yudin}\ \emph {et~al.}(2006)\citenamefont {Yudin},
  \citenamefont {Bandrauk},\ and\ \citenamefont {Corkum}}]{Yudin2006}%
  \BibitemOpen
  \bibfield  {author} {\bibinfo {author} {\bibfnamefont {G.~L.}\ \bibnamefont
  {Yudin}}, \bibinfo {author} {\bibfnamefont {A.~D.}\ \bibnamefont {Bandrauk}},
  \ and\ \bibinfo {author} {\bibfnamefont {P.~B.}\ \bibnamefont {Corkum}},\
  }\href {\doibase 10.1103/PhysRevLett.96.063002} {\bibfield  {journal}
  {\bibinfo  {journal} {Phys. Rev. Lett.}\ }\textbf {\bibinfo {volume} {96}},\
  \bibinfo {pages} {063002} (\bibinfo {year} {2006})}\BibitemShut {NoStop}%
\bibitem [{\citenamefont {Loh}\ \emph {et~al.}(2007)\citenamefont {Loh},
  \citenamefont {Khalil}, \citenamefont {Correa} \emph {et~al.}}]{loh2007}%
  \BibitemOpen
  \bibfield  {author} {\bibinfo {author} {\bibfnamefont {Z.-H.}\ \bibnamefont
  {Loh}}, \bibinfo {author} {\bibfnamefont {M.}~\bibnamefont {Khalil}},
  \bibinfo {author} {\bibfnamefont {R.~E.}\ \bibnamefont {Correa}},  \emph
  {et~al.},\ }\href@noop {} {\bibfield  {journal} {\bibinfo  {journal} {Phys.
  Rev. Lett.}\ }\textbf {\bibinfo {volume} {98}},\ \bibinfo {pages} {143601}
  (\bibinfo {year} {2007})}\BibitemShut {NoStop}%
\bibitem [{\citenamefont {Holler}\ \emph {et~al.}(2011)\citenamefont {Holler},
  \citenamefont {Schapper}, \citenamefont {Gallmann} \emph
  {et~al.}}]{holler2011}%
  \BibitemOpen
  \bibfield  {author} {\bibinfo {author} {\bibfnamefont {M.}~\bibnamefont
  {Holler}}, \bibinfo {author} {\bibfnamefont {F.}~\bibnamefont {Schapper}},
  \bibinfo {author} {\bibfnamefont {L.}~\bibnamefont {Gallmann}},  \emph
  {et~al.},\ }\href@noop {} {\bibfield  {journal} {\bibinfo  {journal} {Phys.
  Rev. Lett.}\ }\textbf {\bibinfo {volume} {106}},\ \bibinfo {pages} {123601}
  (\bibinfo {year} {2011})}\BibitemShut {NoStop}%
\bibitem [{\citenamefont {Bredtmann}\ \emph {et~al.}(2011)\citenamefont
  {Bredtmann}, \citenamefont {Chelkowski},\ and\ \citenamefont
  {Bandrauk}}]{Bredtmann2011}%
  \BibitemOpen
  \bibfield  {author} {\bibinfo {author} {\bibfnamefont {T.}~\bibnamefont
  {Bredtmann}}, \bibinfo {author} {\bibfnamefont {S.}~\bibnamefont
  {Chelkowski}}, \ and\ \bibinfo {author} {\bibfnamefont {A.~D.}\ \bibnamefont
  {Bandrauk}},\ }\href@noop {} {\bibfield  {journal} {\bibinfo  {journal}
  {Phys. Rev. A}\ }\textbf {\bibinfo {volume} {84}},\ \bibinfo {pages}
  {021401(R)} (\bibinfo {year} {2011})}\BibitemShut {NoStop}%
\bibitem [{\citenamefont {Kraus}\ \emph {et~al.}(2013)\citenamefont {Kraus},
  \citenamefont {Zhang}, \citenamefont {Gijsbertsen} \emph
  {et~al.}}]{kraus2013}%
  \BibitemOpen
  \bibfield  {author} {\bibinfo {author} {\bibfnamefont {P.~M.}\ \bibnamefont
  {Kraus}}, \bibinfo {author} {\bibfnamefont {S.~B.}\ \bibnamefont {Zhang}},
  \bibinfo {author} {\bibfnamefont {A.}~\bibnamefont {Gijsbertsen}},  \emph
  {et~al.},\ }\href@noop {} {\bibfield  {journal} {\bibinfo  {journal} {Phys.
  Rev. Lett.}\ }\textbf {\bibinfo {volume} {111}},\ \bibinfo {pages} {243005}
  (\bibinfo {year} {2013})}\BibitemShut {NoStop}%
\bibitem [{\citenamefont {Liu}\ \emph {et~al.}(2014)\citenamefont {Liu},
  \citenamefont {Zeng}, \citenamefont {Li} \emph {et~al.}}]{liu2014}%
  \BibitemOpen
  \bibfield  {author} {\bibinfo {author} {\bibfnamefont {C.}~\bibnamefont
  {Liu}}, \bibinfo {author} {\bibfnamefont {Z.}~\bibnamefont {Zeng}}, \bibinfo
  {author} {\bibfnamefont {R.}~\bibnamefont {Li}},  \emph {et~al.},\
  }\href@noop {} {\bibfield  {journal} {\bibinfo  {journal} {Phys. Rev. A}\
  }\textbf {\bibinfo {volume} {90}},\ \bibinfo {pages} {013403} (\bibinfo
  {year} {2014})}\BibitemShut {NoStop}%
\bibitem [{\citenamefont {Liu}\ \emph {et~al.}(2015)\citenamefont {Liu},
  \citenamefont {Cavaletto}, \citenamefont {Ott}, \citenamefont {Meyer},
  \citenamefont {Mi}, \citenamefont {Harman}, \citenamefont {Keitel},\ and\
  \citenamefont {Pfeifer}}]{liu2015}%
  \BibitemOpen
  \bibfield  {author} {\bibinfo {author} {\bibfnamefont {Z.}~\bibnamefont
  {Liu}}, \bibinfo {author} {\bibfnamefont {S.~M.}\ \bibnamefont {Cavaletto}},
  \bibinfo {author} {\bibfnamefont {C.}~\bibnamefont {Ott}}, \bibinfo {author}
  {\bibfnamefont {K.}~\bibnamefont {Meyer}}, \bibinfo {author} {\bibfnamefont
  {Y.}~\bibnamefont {Mi}}, \bibinfo {author} {\bibfnamefont {Z.}~\bibnamefont
  {Harman}}, \bibinfo {author} {\bibfnamefont {C.~H.}\ \bibnamefont {Keitel}},
  \ and\ \bibinfo {author} {\bibfnamefont {T.}~\bibnamefont {Pfeifer}},\ }\href
  {\doibase 10.1103/PhysRevLett.115.033003} {\bibfield  {journal} {\bibinfo
  {journal} {Phys. Rev. Lett.}\ }\textbf {\bibinfo {volume} {115}},\ \bibinfo
  {pages} {033003} (\bibinfo {year} {2015})}\BibitemShut {NoStop}%
\bibitem [{\citenamefont {Loh}\ and\ \citenamefont {Leone}(2008)}]{Loh2008}%
  \BibitemOpen
  \bibfield  {author} {\bibinfo {author} {\bibfnamefont {Z.-H.}\ \bibnamefont
  {Loh}}\ and\ \bibinfo {author} {\bibfnamefont {S.~R.}\ \bibnamefont
  {Leone}},\ }\href@noop {} {\bibfield  {journal} {\bibinfo  {journal} {J.
  Chem. Phys.}\ }\textbf {\bibinfo {volume} {128}},\ \bibinfo {pages} {204302}
  (\bibinfo {year} {2008})}\BibitemShut {NoStop}%
\bibitem [{\citenamefont {Goulielmakis}\ \emph {et~al.}(2010)\citenamefont
  {Goulielmakis}, \citenamefont {Loh}, \citenamefont {Wirth} \emph
  {et~al.}}]{Goulielmakis2010}%
  \BibitemOpen
  \bibfield  {author} {\bibinfo {author} {\bibfnamefont {E.}~\bibnamefont
  {Goulielmakis}}, \bibinfo {author} {\bibfnamefont {Z.~H.}\ \bibnamefont
  {Loh}}, \bibinfo {author} {\bibfnamefont {A.}~\bibnamefont {Wirth}},  \emph
  {et~al.},\ }\href@noop {} {\bibfield  {journal} {\bibinfo  {journal}
  {Nature}\ }\textbf {\bibinfo {volume} {466}},\ \bibinfo {pages} {739}
  (\bibinfo {year} {2010})}\BibitemShut {NoStop}%
\bibitem [{\citenamefont {Ott}\ \emph {et~al.}(2013)\citenamefont {Ott},
  \citenamefont {Kaldun}, \citenamefont {Raith} \emph {et~al.}}]{ott2013}%
  \BibitemOpen
  \bibfield  {author} {\bibinfo {author} {\bibfnamefont {C.}~\bibnamefont
  {Ott}}, \bibinfo {author} {\bibfnamefont {A.}~\bibnamefont {Kaldun}},
  \bibinfo {author} {\bibfnamefont {P.}~\bibnamefont {Raith}},  \emph
  {et~al.},\ }\href@noop {} {\bibfield  {journal} {\bibinfo  {journal}
  {Science}\ }\textbf {\bibinfo {volume} {340}},\ \bibinfo {pages} {716}
  (\bibinfo {year} {2013})}\BibitemShut {NoStop}%
\bibitem [{\citenamefont {Chini}\ \emph {et~al.}(2013)\citenamefont {Chini},
  \citenamefont {Wang}, \citenamefont {Cheng} \emph {et~al.}}]{Chini2013}%
  \BibitemOpen
  \bibfield  {author} {\bibinfo {author} {\bibfnamefont {M.}~\bibnamefont
  {Chini}}, \bibinfo {author} {\bibfnamefont {X.}~\bibnamefont {Wang}},
  \bibinfo {author} {\bibfnamefont {Y.}~\bibnamefont {Cheng}},  \emph
  {et~al.},\ }\href@noop {} {\bibfield  {journal} {\bibinfo  {journal} {Sci.
  Rep.}\ }\textbf {\bibinfo {volume} {3}},\ \bibinfo {pages} {1105} (\bibinfo
  {year} {2013})}\BibitemShut {NoStop}%
\bibitem [{\citenamefont {Kaldun}\ \emph {et~al.}(2014)\citenamefont {Kaldun},
  \citenamefont {Ott}, \citenamefont {Blättermann} \emph
  {et~al.}}]{kaldun2014}%
  \BibitemOpen
  \bibfield  {author} {\bibinfo {author} {\bibfnamefont {A.}~\bibnamefont
  {Kaldun}}, \bibinfo {author} {\bibfnamefont {C.}~\bibnamefont {Ott}},
  \bibinfo {author} {\bibfnamefont {A.}~\bibnamefont {Blättermann}},  \emph
  {et~al.},\ }\href@noop {} {\bibfield  {journal} {\bibinfo  {journal} {Phys.
  Rev. Lett.}\ }\textbf {\bibinfo {volume} {112}},\ \bibinfo {pages} {103001}
  (\bibinfo {year} {2014})}\BibitemShut {NoStop}%
\bibitem [{\citenamefont {Schweigert}\ and\ \citenamefont
  {Mukamel}(2007)}]{schweigert2007}%
  \BibitemOpen
  \bibfield  {author} {\bibinfo {author} {\bibfnamefont {I.~V.}\ \bibnamefont
  {Schweigert}}\ and\ \bibinfo {author} {\bibfnamefont {S.}~\bibnamefont
  {Mukamel}},\ }\href@noop {} {\bibfield  {journal} {\bibinfo  {journal} {Phys.
  Rev. A}\ }\textbf {\bibinfo {volume} {76}},\ \bibinfo {pages} {012504}
  (\bibinfo {year} {2007})}\BibitemShut {NoStop}%
\bibitem [{\citenamefont {Harbola}\ and\ \citenamefont
  {Mukamel}(2009)}]{harbola2009}%
  \BibitemOpen
  \bibfield  {author} {\bibinfo {author} {\bibfnamefont {U.}~\bibnamefont
  {Harbola}}\ and\ \bibinfo {author} {\bibfnamefont {S.}~\bibnamefont
  {Mukamel}},\ }\href@noop {} {\bibfield  {journal} {\bibinfo  {journal} {Phys.
  Rev. B}\ }\textbf {\bibinfo {volume} {79}},\ \bibinfo {pages} {085108}
  (\bibinfo {year} {2009})}\BibitemShut {NoStop}%
\bibitem [{\citenamefont {Rohringer}\ \emph {et~al.}(2015)\citenamefont
  {Rohringer}, \citenamefont {Kimberg}, \citenamefont {Weninger}, \citenamefont
  {Sanchez-Gonzalez}, \citenamefont {Lutman}, \citenamefont {Maxwell},
  \citenamefont {Bostedt}, \citenamefont {Monterro} \emph
  {et~al.}}]{Rohringe2015}%
  \BibitemOpen
  \bibfield  {author} {\bibinfo {author} {\bibfnamefont {N.}~\bibnamefont
  {Rohringer}}, \bibinfo {author} {\bibfnamefont {V.}~\bibnamefont {Kimberg}},
  \bibinfo {author} {\bibfnamefont {C.}~\bibnamefont {Weninger}}, \bibinfo
  {author} {\bibfnamefont {A.}~\bibnamefont {Sanchez-Gonzalez}}, \bibinfo
  {author} {\bibfnamefont {A.}~\bibnamefont {Lutman}}, \bibinfo {author}
  {\bibfnamefont {T.}~\bibnamefont {Maxwell}}, \bibinfo {author} {\bibfnamefont
  {C.}~\bibnamefont {Bostedt}}, \bibinfo {author} {\bibfnamefont {S.~C.}\
  \bibnamefont {Monterro}},  \emph {et~al.},\ }in\ \href@noop {} {\emph
  {\bibinfo {booktitle} {X-Ray Lasers 2014}}}\ (\bibinfo  {publisher}
  {Springer},\ \bibinfo {year} {2015})\ p.\ \bibinfo {pages} {In
  press}\BibitemShut {NoStop}%
\bibitem [{\citenamefont {Pahl}\ \emph {et~al.}(1996)\citenamefont {Pahl},
  \citenamefont {Meyer},\ and\ \citenamefont {Cederbaum}}]{Pahl1996}%
  \BibitemOpen
  \bibfield  {author} {\bibinfo {author} {\bibfnamefont {E.}~\bibnamefont
  {Pahl}}, \bibinfo {author} {\bibfnamefont {H.~D.}\ \bibnamefont {Meyer}}, \
  and\ \bibinfo {author} {\bibfnamefont {L.~S.}\ \bibnamefont {Cederbaum}},\
  }\href@noop {} {\bibfield  {journal} {\bibinfo  {journal} {Z. Phys. D-Atoms
  Mol. Clusters}\ }\textbf {\bibinfo {volume} {38}},\ \bibinfo {pages} {215}
  (\bibinfo {year} {1996})}\BibitemShut {NoStop}%
\bibitem [{\citenamefont {Zhang}\ and\ \citenamefont
  {Rohringer}(2014)}]{Zhang2014}%
  \BibitemOpen
  \bibfield  {author} {\bibinfo {author} {\bibfnamefont {S.~B.}\ \bibnamefont
  {Zhang}}\ and\ \bibinfo {author} {\bibfnamefont {N.}~\bibnamefont
  {Rohringer}},\ }\href@noop {} {\bibfield  {journal} {\bibinfo  {journal}
  {Phys. Rev. A}\ }\textbf {\bibinfo {volume} {89}},\ \bibinfo {pages} {013407}
  (\bibinfo {year} {2014})}\BibitemShut {NoStop}%
\bibitem [{\citenamefont {Gamaly}(2011)}]{Gamaly2011}%
  \BibitemOpen
  \bibfield  {author} {\bibinfo {author} {\bibfnamefont {E.}~\bibnamefont
  {Gamaly}},\ }\href@noop {} {\emph {\bibinfo {title} {Femtosecond laser-matter
  Interaction: Theory, Experiments and Applications}}}\ (\bibinfo  {publisher}
  {Pan Stanford Publishing Pte. Ltd.},\ \bibinfo {address} {Singapore},\
  \bibinfo {year} {2011})\BibitemShut {NoStop}%
\bibitem [{\citenamefont {Shore}(2011)}]{Shore2011}%
  \BibitemOpen
  \bibfield  {author} {\bibinfo {author} {\bibfnamefont {B.~W.}\ \bibnamefont
  {Shore}},\ }\href@noop {} {\emph {\bibinfo {title} {Manipulating Quantum
  Structures Using Laser Pulses}}}\ (\bibinfo  {publisher} {Cambridge
  University Press},\ \bibinfo {address} {New York},\ \bibinfo {year}
  {2011})\BibitemShut {NoStop}%
\bibitem [{\citenamefont {Cederbaum}\ and\ \citenamefont
  {Domcke}(1981)}]{Cederbaum1981}%
  \BibitemOpen
  \bibfield  {author} {\bibinfo {author} {\bibfnamefont {L.~S.}\ \bibnamefont
  {Cederbaum}}\ and\ \bibinfo {author} {\bibfnamefont {W.}~\bibnamefont
  {Domcke}},\ }\href@noop {} {\bibfield  {journal} {\bibinfo  {journal} {J.
  Phys. B}\ }\textbf {\bibinfo {volume} {14}},\ \bibinfo {pages} {4665}
  (\bibinfo {year} {1981})}\BibitemShut {NoStop}%
\bibitem [{\citenamefont {Domcke}(1991)}]{Domcke1991}%
  \BibitemOpen
  \bibfield  {author} {\bibinfo {author} {\bibfnamefont {W.}~\bibnamefont
  {Domcke}},\ }\href@noop {} {\bibfield  {journal} {\bibinfo  {journal} {Phys.
  Rep.}\ }\textbf {\bibinfo {volume} {208}},\ \bibinfo {pages} {97} (\bibinfo
  {year} {1991})}\BibitemShut {NoStop}%
\bibitem [{\citenamefont {Gel'mukhanov}\ and\ \citenamefont
  {{\AA}gren}(1999)}]{gel1999}%
  \BibitemOpen
  \bibfield  {author} {\bibinfo {author} {\bibfnamefont {F.}~\bibnamefont
  {Gel'mukhanov}}\ and\ \bibinfo {author} {\bibfnamefont {H.}~\bibnamefont
  {{\AA}gren}},\ }\href@noop {} {\bibfield  {journal} {\bibinfo  {journal}
  {Physics Reports}\ }\textbf {\bibinfo {volume} {312}},\ \bibinfo {pages} {87}
  (\bibinfo {year} {1999})}\BibitemShut {NoStop}%
\bibitem [{\citenamefont {Skytt}\ \emph {et~al.}(1997)\citenamefont {Skytt},
  \citenamefont {Glans}, \citenamefont {Gunnelin} \emph {et~al.}}]{Skytt1997}%
  \BibitemOpen
  \bibfield  {author} {\bibinfo {author} {\bibfnamefont {P.}~\bibnamefont
  {Skytt}}, \bibinfo {author} {\bibfnamefont {P.}~\bibnamefont {Glans}},
  \bibinfo {author} {\bibfnamefont {K.}~\bibnamefont {Gunnelin}},  \emph
  {et~al.},\ }\href@noop {} {\bibfield  {journal} {\bibinfo  {journal} {Phys.
  Rev. A}\ }\textbf {\bibinfo {volume} {55}},\ \bibinfo {pages} {134} (\bibinfo
  {year} {1997})}\BibitemShut {NoStop}%
\bibitem [{\citenamefont {Hack}\ and\ \citenamefont {Huber}(1991)}]{hack1991}%
  \BibitemOpen
  \bibfield  {author} {\bibinfo {author} {\bibfnamefont {E.}~\bibnamefont
  {Hack}}\ and\ \bibinfo {author} {\bibfnamefont {J.}~\bibnamefont {Huber}},\
  }\href@noop {} {\bibfield  {journal} {\bibinfo  {journal} {Int. Rev. Phys.
  Chem.}\ }\textbf {\bibinfo {volume} {10}},\ \bibinfo {pages} {287} (\bibinfo
  {year} {1991})}\BibitemShut {NoStop}%
\bibitem [{\citenamefont {Carter}\ and\ \citenamefont
  {Huber}(2000)}]{carter2000}%
  \BibitemOpen
  \bibfield  {author} {\bibinfo {author} {\bibfnamefont {R.~T.}\ \bibnamefont
  {Carter}}\ and\ \bibinfo {author} {\bibfnamefont {J.~R.}\ \bibnamefont
  {Huber}},\ }\href@noop {} {\bibfield  {journal} {\bibinfo  {journal} {Chem.
  Soc. Rev.}\ }\textbf {\bibinfo {volume} {29}},\ \bibinfo {pages} {305}
  (\bibinfo {year} {2000})}\BibitemShut {NoStop}%
\bibitem [{\citenamefont {Lutman}\ \emph {et~al.}(2013)\citenamefont {Lutman},
  \citenamefont {Coffee}, \citenamefont {Ding}, \citenamefont {Huang},
  \citenamefont {Krzywinski}, \citenamefont {Maxwell}, \citenamefont
  {Messerschmidt},\ and\ \citenamefont {Nuhn}}]{lutman2013}%
  \BibitemOpen
  \bibfield  {author} {\bibinfo {author} {\bibfnamefont {A.}~\bibnamefont
  {Lutman}}, \bibinfo {author} {\bibfnamefont {R.}~\bibnamefont {Coffee}},
  \bibinfo {author} {\bibfnamefont {Y.}~\bibnamefont {Ding}}, \bibinfo {author}
  {\bibfnamefont {Z.}~\bibnamefont {Huang}}, \bibinfo {author} {\bibfnamefont
  {J.}~\bibnamefont {Krzywinski}}, \bibinfo {author} {\bibfnamefont
  {T.}~\bibnamefont {Maxwell}}, \bibinfo {author} {\bibfnamefont
  {M.}~\bibnamefont {Messerschmidt}}, \ and\ \bibinfo {author} {\bibfnamefont
  {H.-D.}\ \bibnamefont {Nuhn}},\ }\href@noop {} {\bibfield  {journal}
  {\bibinfo  {journal} {Phys. Rev. Lett.}\ }\textbf {\bibinfo {volume} {110}},\
  \bibinfo {pages} {134801} (\bibinfo {year} {2013})}\BibitemShut {NoStop}%
\bibitem [{\citenamefont {Allaria}\ \emph {et~al.}(2013)\citenamefont
  {Allaria}, \citenamefont {Bencivenga}, \citenamefont {Borghes}, \citenamefont
  {Capotondi}, \citenamefont {Castronovo}, \citenamefont {Charalambous},
  \citenamefont {Cinquegrana}, \citenamefont {Danailov}, \citenamefont
  {De~Ninno}, \citenamefont {Demidovich} \emph {et~al.}}]{allaria2013}%
  \BibitemOpen
  \bibfield  {author} {\bibinfo {author} {\bibfnamefont {E.}~\bibnamefont
  {Allaria}}, \bibinfo {author} {\bibfnamefont {F.}~\bibnamefont {Bencivenga}},
  \bibinfo {author} {\bibfnamefont {R.}~\bibnamefont {Borghes}}, \bibinfo
  {author} {\bibfnamefont {F.}~\bibnamefont {Capotondi}}, \bibinfo {author}
  {\bibfnamefont {D.}~\bibnamefont {Castronovo}}, \bibinfo {author}
  {\bibfnamefont {P.}~\bibnamefont {Charalambous}}, \bibinfo {author}
  {\bibfnamefont {P.}~\bibnamefont {Cinquegrana}}, \bibinfo {author}
  {\bibfnamefont {M.}~\bibnamefont {Danailov}}, \bibinfo {author}
  {\bibfnamefont {G.}~\bibnamefont {De~Ninno}}, \bibinfo {author}
  {\bibfnamefont {A.}~\bibnamefont {Demidovich}},  \emph {et~al.},\ }\href@noop
  {} {\bibfield  {journal} {\bibinfo  {journal} {Nat. Commun.}\ }\textbf
  {\bibinfo {volume} {4}},\ \bibinfo {pages} {2476} (\bibinfo {year}
  {2013})}\BibitemShut {NoStop}%
\bibitem [{\citenamefont {Marinelli}\ \emph {et~al.}(2015)\citenamefont
  {Marinelli}, \citenamefont {Ratner}, \citenamefont {Lutman}, \citenamefont
  {Turner}, \citenamefont {Welch}, \citenamefont {Decker}, \citenamefont
  {Loos}, \citenamefont {Behrens}, \citenamefont {Gilevich}, \citenamefont
  {Miahnahri} \emph {et~al.}}]{marinelli2015}%
  \BibitemOpen
  \bibfield  {author} {\bibinfo {author} {\bibfnamefont {A.}~\bibnamefont
  {Marinelli}}, \bibinfo {author} {\bibfnamefont {D.}~\bibnamefont {Ratner}},
  \bibinfo {author} {\bibfnamefont {A.}~\bibnamefont {Lutman}}, \bibinfo
  {author} {\bibfnamefont {J.}~\bibnamefont {Turner}}, \bibinfo {author}
  {\bibfnamefont {J.}~\bibnamefont {Welch}}, \bibinfo {author} {\bibfnamefont
  {F.-J.}\ \bibnamefont {Decker}}, \bibinfo {author} {\bibfnamefont
  {H.}~\bibnamefont {Loos}}, \bibinfo {author} {\bibfnamefont {C.}~\bibnamefont
  {Behrens}}, \bibinfo {author} {\bibfnamefont {S.}~\bibnamefont {Gilevich}},
  \bibinfo {author} {\bibfnamefont {A.}~\bibnamefont {Miahnahri}},  \emph
  {et~al.},\ }\href@noop {} {\bibfield  {journal} {\bibinfo  {journal} {Nat.
  Commun.}\ }\textbf {\bibinfo {volume} {6}},\ \bibinfo {pages} {6369}
  (\bibinfo {year} {2015})}\BibitemShut {NoStop}%
\bibitem [{\citenamefont {Hartmann}\ \emph {et~al.}(2014)\citenamefont
  {Hartmann}, \citenamefont {Helml}, \citenamefont {Galler}, \citenamefont
  {Bionta}, \citenamefont {Gr{\"u}nert}, \citenamefont {Molodtsov},
  \citenamefont {Ferguson}, \citenamefont {Schorb}, \citenamefont {Swiggers},
  \citenamefont {Carron} \emph {et~al.}}]{hartmann2014}%
  \BibitemOpen
  \bibfield  {author} {\bibinfo {author} {\bibfnamefont {N.}~\bibnamefont
  {Hartmann}}, \bibinfo {author} {\bibfnamefont {W.}~\bibnamefont {Helml}},
  \bibinfo {author} {\bibfnamefont {A.}~\bibnamefont {Galler}}, \bibinfo
  {author} {\bibfnamefont {M.}~\bibnamefont {Bionta}}, \bibinfo {author}
  {\bibfnamefont {J.}~\bibnamefont {Gr{\"u}nert}}, \bibinfo {author}
  {\bibfnamefont {S.}~\bibnamefont {Molodtsov}}, \bibinfo {author}
  {\bibfnamefont {K.}~\bibnamefont {Ferguson}}, \bibinfo {author}
  {\bibfnamefont {S.}~\bibnamefont {Schorb}}, \bibinfo {author} {\bibfnamefont
  {M.}~\bibnamefont {Swiggers}}, \bibinfo {author} {\bibfnamefont
  {S.}~\bibnamefont {Carron}},  \emph {et~al.},\ }\href@noop {} {\bibfield
  {journal} {\bibinfo  {journal} {Nat. Photon.}\ }\textbf {\bibinfo {volume}
  {8}},\ \bibinfo {pages} {706} (\bibinfo {year} {2014})}\BibitemShut {NoStop}%
\bibitem [{\citenamefont {Shankar}(2012)}]{shankar2012}%
  \BibitemOpen
  \bibfield  {author} {\bibinfo {author} {\bibfnamefont {R.}~\bibnamefont
  {Shankar}},\ }\href@noop {} {\emph {\bibinfo {title} {Principles of quantum
  mechanics}}}\ (\bibinfo  {publisher} {Springer Science \& Business Media},\
  \bibinfo {year} {2012})\BibitemShut {NoStop}%
\end{thebibliography}%

\onecolumngrid

\begin{figure}[htbf]
\includegraphics[width= 8.0 cm]{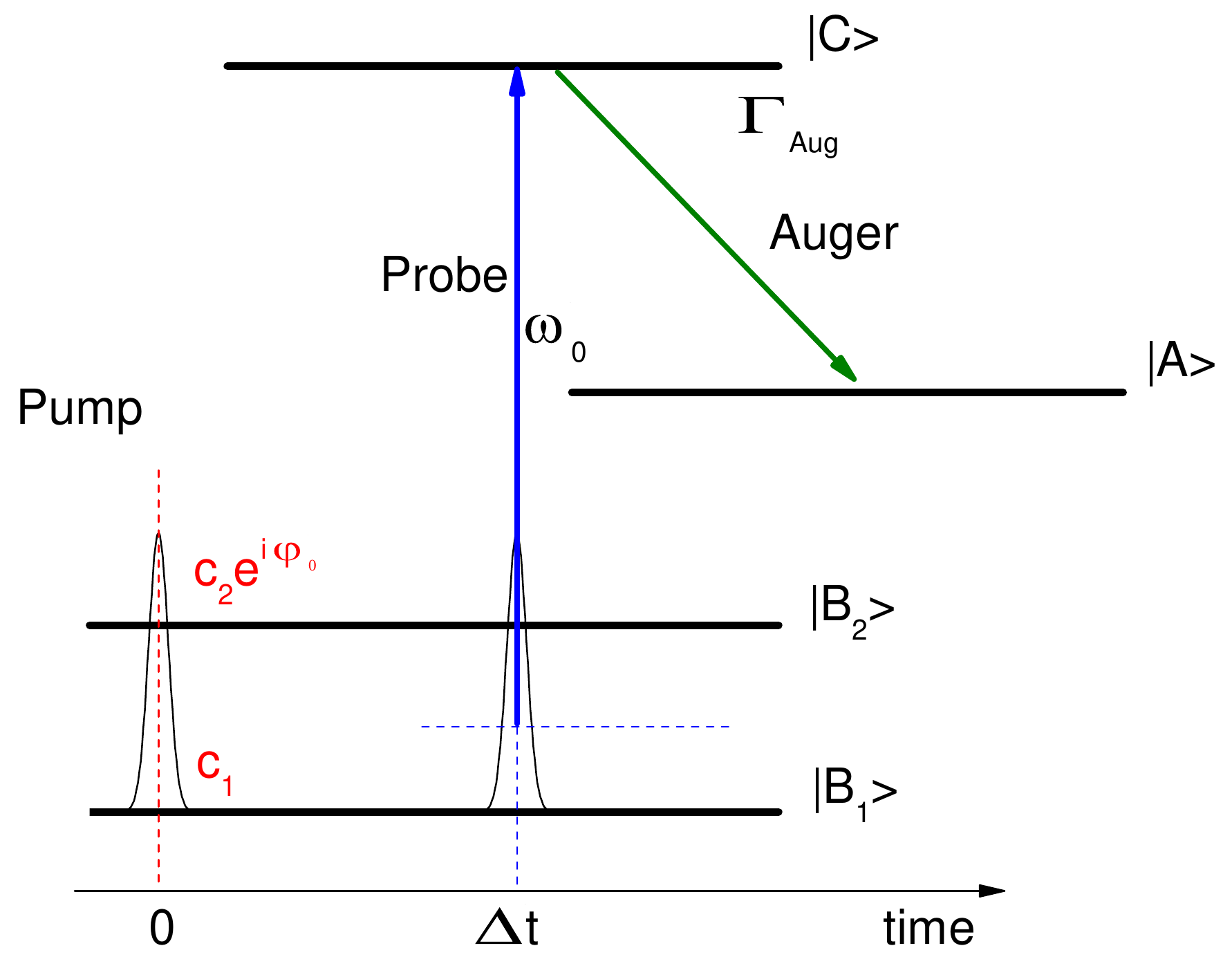}
\caption{(Color online) Schematics of the Auger process triggered by the weak short x-ray probe pulse from a two-state coherent wave packet with initial populations $c_1^2$ and $c_2^2$ and relative phase $\varphi_0$. The coherent wave packet is supposed to be created by an appropriate pump pulse and the probe pulse is time delayed by $\Delta t$.}
\end{figure}

\begin{figure}[htbf]
\includegraphics[width= 16.0 cm]{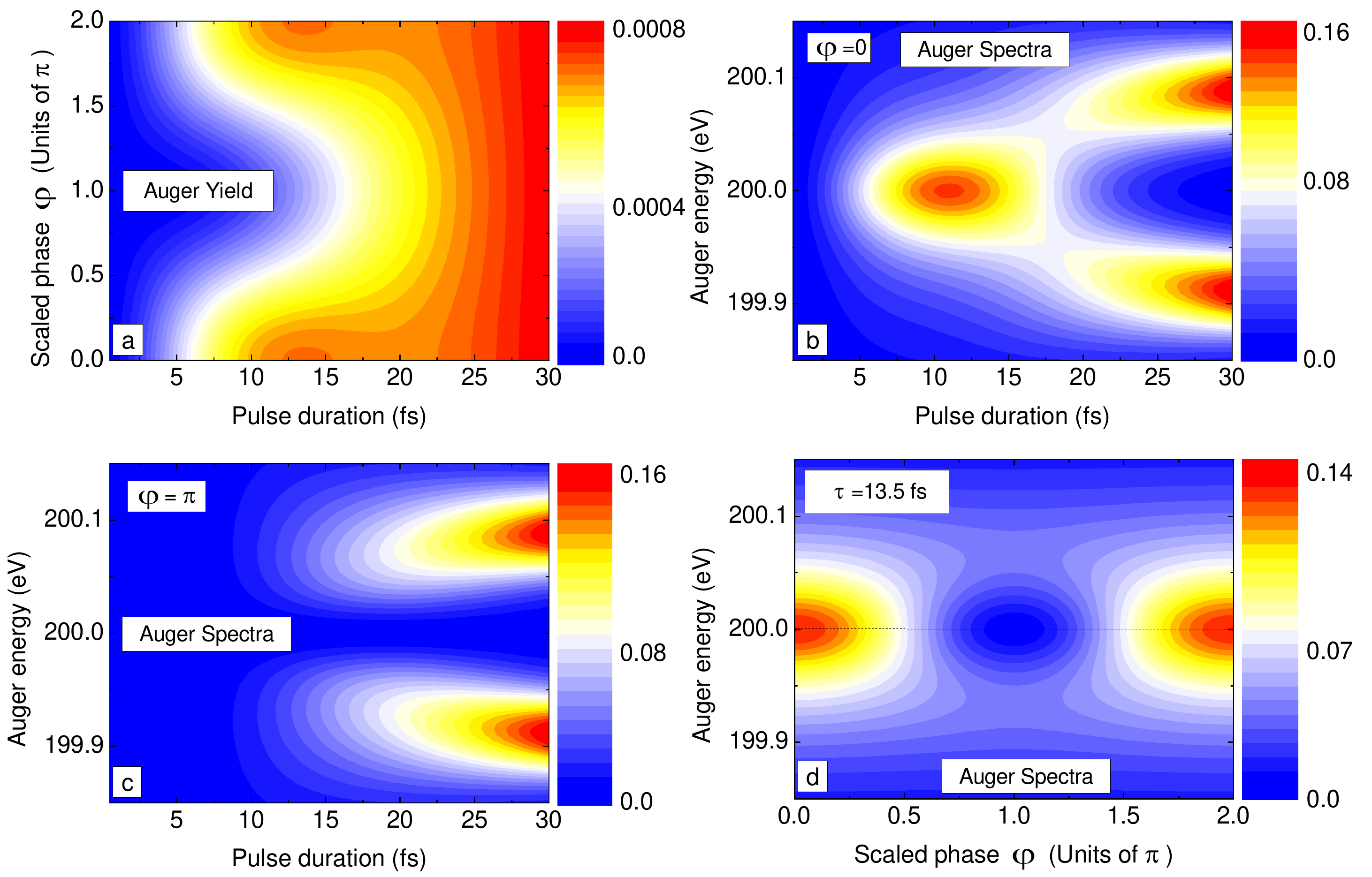}
\caption{(Color online) The total Auger yield and Auger electron spectra for the case with initial state populations $c_1^2=c_2^2=0.5$. Note that $E_{B_2B_1}=0.2$ eV, $\omega_0=E_{CB_1}-E_{B_2B_1}/2=209.9$ eV and $\Omega_1=\Omega_2=0.0001$ a.u.. Explicitly, panel (a) shows the total Auger yield with respect to the pulse duration $\tau$ and the scaled phase $\varphi$; panels (b) and (c) show the Auger electron spectra with respect to the pulse duration for the scaled phase $\varphi=0$ and $\varphi=\pi$, respectively; panel (d) shows the Auger electron spectra with respect to the scaled phase $\varphi$ for the pulse with fixed pulse duration $\tau=13.5$ fs. The scaled phase is defined as $\varphi=mod(\varphi_0-E_{B_2B_1}\Delta t,2\pi)$.}
\end{figure}

\begin{figure}[htbf]
\includegraphics[width= 8.0 cm]{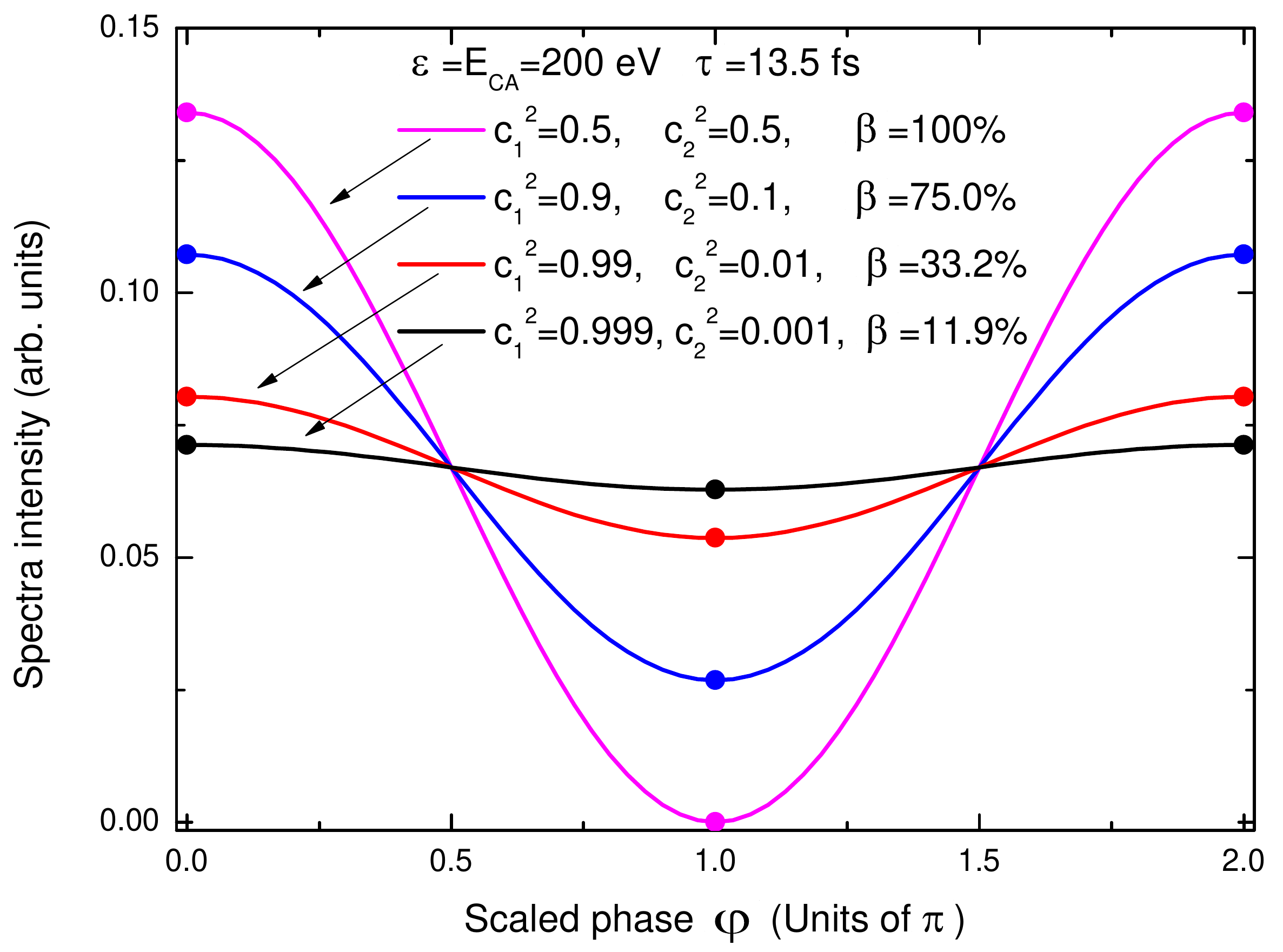}
\caption{(Color online) The scald phase $\varphi$ dependent Auger electron spectra by the pulse with pulse duration $\tau=13.5$ fs  at the Auger energy $\varepsilon=E_{CA}=200$ eV for the cases with different initial state populations. Note that $E_{B_2B_1}=0.2$ eV, $\omega_0=E_{CB_1}-E_{B_2B_1}/2=209.9$ eV and $\Omega_1=\Omega_2=0.0001$ a.u.. $\beta$ is the modulation depth of the spectra, defined as (max-min)/max.}
\end{figure}

\begin{figure}[htbf]
\includegraphics[width= 16.0 cm]{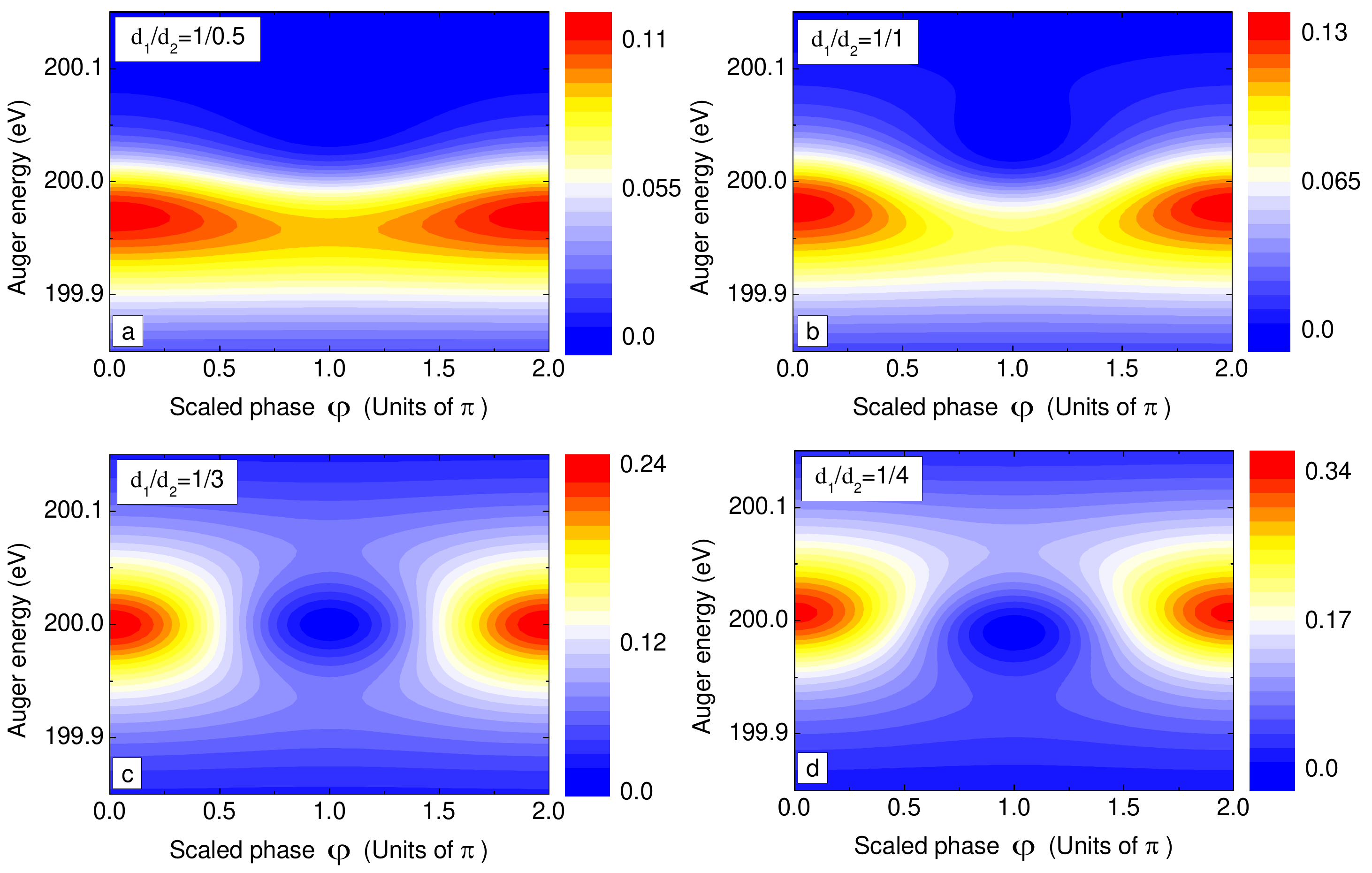}
\caption{(Color online) The scald phase $\varphi$ dependent Auger electron spectra by the pulse with pulse duration $\tau=13.5$ fs for the cases with initial state populations $c_1^2=0.9$ (or $c_2^2=0.1$) and different dipole ratios $\frac{d_1}{d_2}$. The energy gap $E_{B_2B_1}=0.2$ eV and the center frequency of the probe pulse $\omega_0=E_{CB_1}-E_{B_2B_1}/2=209.9$ eV. Panels (a) (b) (c) and (d) correspond to the ratios of $\frac{c_1d_1}{c_2d_2}=\frac{3}{0.5}$, $\frac{3}{1}$, $\frac{3}{3}$ and $\frac{3}{4}$, respectively.}
\end{figure}

\begin{figure}[htbf]
\includegraphics[width= 16.0 cm]{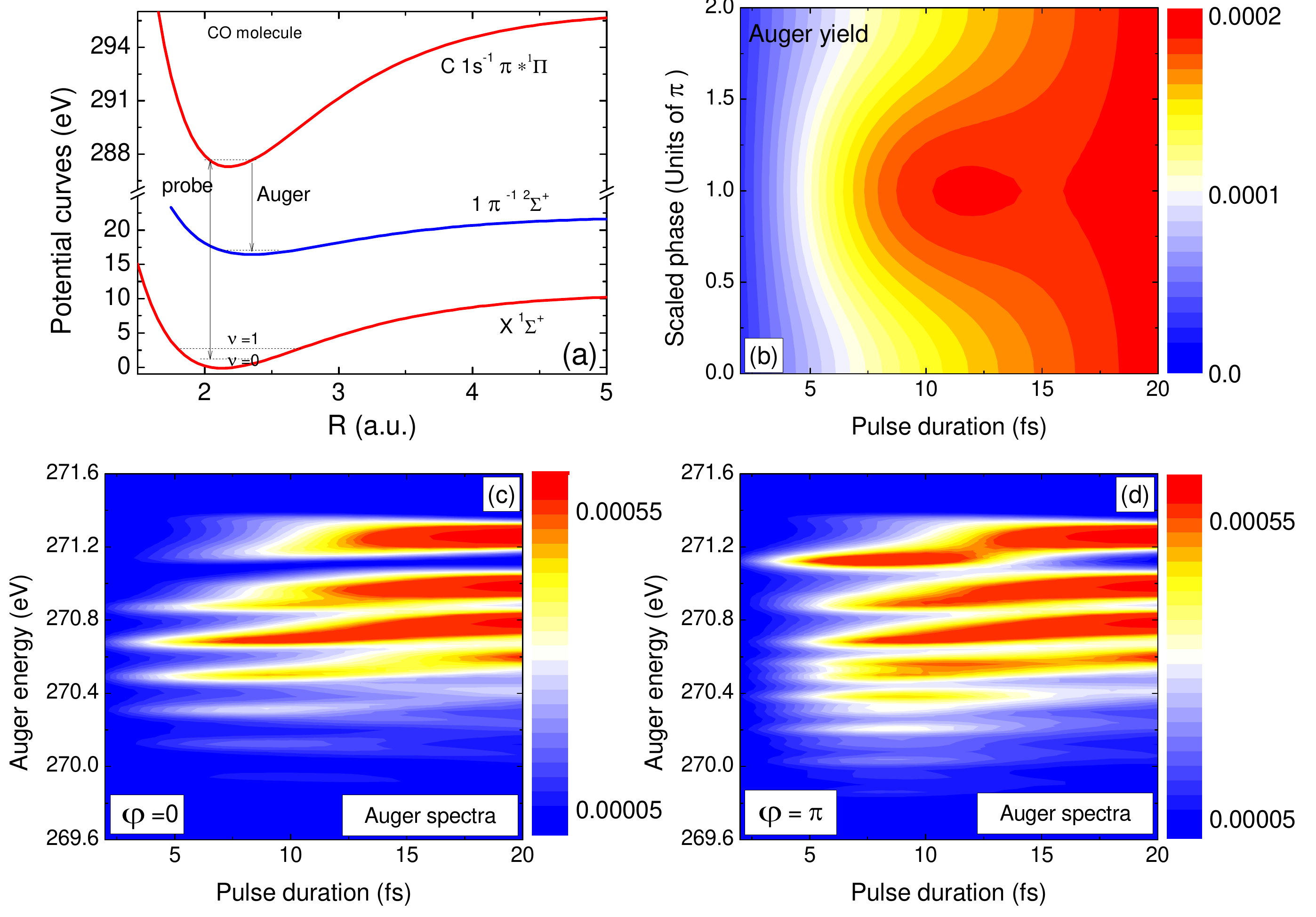}
\caption{(Color online) panel (a) potential curves involved in the study. The vibrational wave packet at time zero is supposed to be $(|\nu=0\rangle+e^{i\phi_0}|\nu=1\rangle)/\sqrt{2}$, and the probe pulse is time delay by $\Delta t$; panel (b) total Auger yield with respect to the pulse duration $\tau$ and the scaled phase $\varphi=mod(\varphi_0-E_{\nu_{01}}\Delta t,2\pi)$; panels (c) and (d) show vibrationally resolved Auger electron spectra with respect to the pulse duration for the scaled phase $\varphi=0$ and $\varphi=\pi$, respectively.}
\end{figure}

\begin{figure}[htbf]
\includegraphics[width= 16.0 cm]{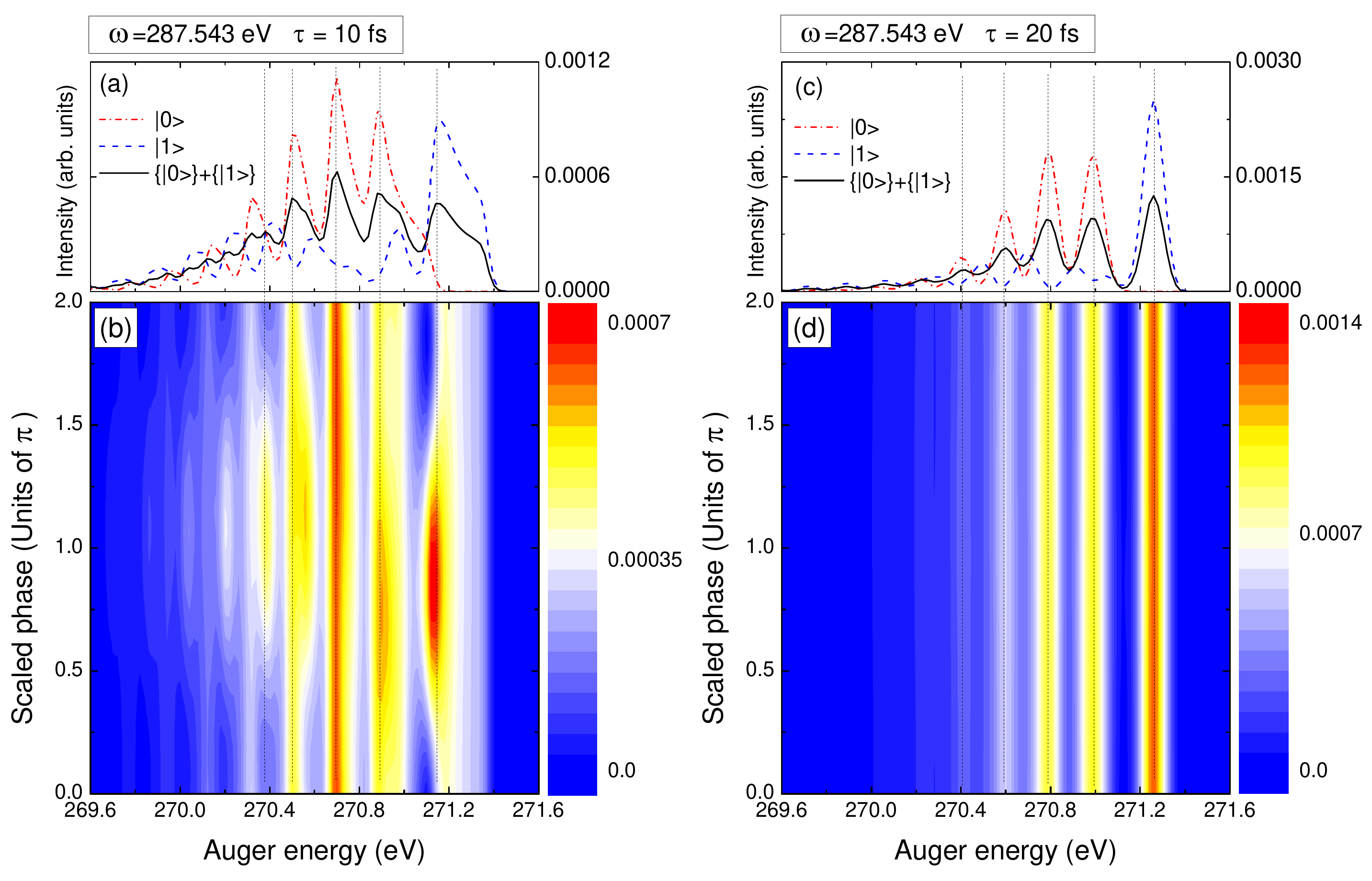}
\caption{(Color online) (a) Auger electron spectra for initial states of vibrational state $|0\rangle$ (red) or $|1\rangle$ (blue) for $\tau$=10 fs and the Auger spectra of an incoherent sum (black)of the initial states $|0\rangle$ and $|1\rangle$ ; (b) vibrationally resolved Auger electron spectra as a function of $\varphi$ for $\tau$=10 fs for the initial coherent wave packet $|0\rangle+|1\rangle$; (c) and (d) are the same as in (a) and (b), respectively, but for $\tau$=20 fs.}
\end{figure}

\end{document}